\documentclass[preprint,showpacs,preprintnumbers,amsmath,amssymb]{revtex4}
\usepackage{graphicx}
\begin{document}
%
\title{Hybrid and Conventional Baryons in the Flux-Tube Model}
%
%
\author{Simon~Capstick}
\email{capstick@csit.fsu.edu}
\affiliation{
	Department of Physics, Florida State University, 
	Tallahassee, FL 32306-4350, USA}
\author{Philip~R.~Page}
\email{prp@lanl.gov}
\affiliation{
	Theoretical Division, MS B283, Los Alamos National Laboratory,
	Los Alamos, NM 87545, USA}
\date{\today}
\begin{abstract}
Conventional and hybrid light quark baryons are constructed in the
non-relativistic flux-tube model of Isgur and Paton, which is
motivated by lattice QCD. The motion of the flux tube with the three
quark positions fixed, except for center of mass corrections, is
discussed. It is shown that the problem can be reduced to the
independent motion of the junction and the strings connecting the
junction to the quarks. The important role played by quark-exchange
symmetry in constraining the flavor structure of (hybrid) baryons is
emphasized.  The flavor, quark spin $S$ and $J^{P}$ of the seven
low-lying hybrid baryons are found to be $(N,\Delta)^{2S+1}J^P =$ $
N^2 {\frac{1}{2}}^+, \; N^2 {\frac{3}{2}}^+, \; \Delta^4
{\frac{1}{2}}^+, \; \Delta^4 {\frac{3}{2}}^+, \; \Delta^4
{\frac{5}{2}}^+$, where the $N^2{\frac{1}{2}}^+$ and
$N^2{\frac{3}{2}}^+$ states are doublets. The motion of the three
quarks in an adiabatic potential derived from the flux-tube dynamics
is considered. A mass of $1870\pm 100$ MeV for the lightest nucleon
hybrids is found by employing a variational method.
\end{abstract}
\pacs{12.38.Lg, 12.39.Mk, 12.39.Pn, 12.40.Yx, 14.20.-c}
\keywords{Suggested keywords}
\maketitle
%
%
\def\slash#1{#1 \hskip -0.5em / }  
\def\rmb#1{{\bf #1}}
\def\lpmb#1{\mbox{\boldmath $#1$}}
\def\nn{\nonumber}
\def\>{\rangle}
\def\<{\langle}
\newcommand{\Eqs}[1]{Eqs.~(\protect\ref{#1})}
\newcommand{\Eq}[1]{Eq.~(\protect\ref{#1})}
\newcommand{\Fig}[1]{Fig.~\protect\ref{#1}}
\newcommand{\Figs}[1]{Figs.~\protect\ref{#1}}
\newcommand{\Sec}[1]{Sec.~\protect\ref{#1}}
\newcommand{\Ref}[1]{Ref.~\protect\cite{#1}}
\newcommand{\Refs}[1]{Refs.~\protect\cite{#1}}
\newcommand{\Tab}[1]{Table~\protect\ref{#1}}
\renewcommand{\-}{\!-\!}
\newcommand{\sfrac}[2]{\mbox{$\textstyle \frac{#1}{#2}$}}
\newcommand{\etahat}{{\mbox{\boldmath $\hat{\eta}$}}}
\def\beq{\begin{equation}}
\def\eeq{\end{equation}}
\newcommand{\btab}{\begin{tabbing}}
\newcommand{\etab}{\end{tabbing}}
\newcommand{\eqntimes}{\mbox{} \times}
\newcommand{\eqnhspace}{\hspace{3em}}
\newcommand{\intereqnvspace}{\vspace{-2.1ex}}
\newcommand{\eqnneghspace}{\hspace{-6ex}}
\newcommand{\beqn}{\begin{equation}}
\newcommand{\eeqn}{\end{equation}}
\newcommand{\barr}[1]{\begin{array}{#1}}
\newcommand{\earr}{\end{array}}
\newcommand{\beqna}{\begin{eqnarray}}
\newcommand{\eeqna}{\end{eqnarray}}
\newcommand{\btablec}{\begin{table} \begin{center}}
\newcommand{\etablec}{\end{center} \end{table}}
\newcommand{\lapprox}{\stackrel{\scriptstyle <}{\scriptstyle \sim}}
\newcommand{\rapprox}{\stackrel{\scriptstyle >}{\scriptstyle \sim}}
\newcommand{\gapproxeq}{\lower.7ex\hbox{$\;\stackrel{\textstyle>}{\sim}\;$}}
\newcommand{\lapproxeq}{\lower.7ex\hbox{$\;\stackrel{\textstyle<}{\sim}\;$}} 
\newtheorem{theorem}{Theorem}
\newcommand{\br}{{\bf r}}
\newcommand{\bl}{{\lpmb l}}
\newcommand{\bk} {{\bf k}}
\newcommand{\bp} {{\bf p}}
\newcommand{\brhd} {{\mbox{\boldmath $\rhd$}}}
\newcommand{\brho} {{\mbox{\boldmath $\rho$}}}
\newcommand{\blambda} {{\mbox{\boldmath $\lambda$}}}
\newcommand{\bnabla} {{\mbox{\boldmath $\nabla$}}}
\newcommand{\abst}{{\noindent \bf Abstract \hspace{0.4cm}}}
\newcommand{\plabel}[1]{\label{#1}}
\newcommand{\meff}{M_{\mbox{\small eff}}}
\newcommand{\beff}{m_{\mbox{\small eff}}}
\newcommand{\vbead}{V_{\mbox{\small J}}}
\newcommand{\req}{{\bf r}_{\mbox{eq}}}
\newcommand{\thetaj}{\theta_J}
\newcommand{\thetarl}{\theta_{\rho\lambda}}
\newcommand{\phirho}{{\phi_{\rho}}}
\newcommand{\mr}{{m_r}}
\newcommand{\rcm}{{{\bf R}_{cm}}}
%
\sloppy
\bigskip

\section{Introduction}
Since all possible good ($J^P$) quantum numbers of baryons can be
described by conventional excitations of three quarks, the description
of hybrid baryons, defined as bound states containing explicit
excitations of the gluon fields of QCD, is necessarily model
dependent. Nevertheless, {\it any} model of QCD bound states which
allows the gluon fields to be dynamical degrees of freedom, as opposed
to simply generating a potential (or surface) in which the quarks
move, will have additional states involving excitations of those
degrees of freedom. A description of the spectrum of hybrid baryons,
the degree of mixing between them and conventional $qqq$ excitations,
and their strong decays will therefore be necessary in order to
describe the results of scattering experiments which involve excited
baryons. For example, such experiments make up the excited baryon
resonance (N*) program at Jefferson Laboratory, where many excited
states of baryons are produced electromagnetically. Hybrid baryons
must play a role in such experiments. In principle their presence
can be detected by finding more states than predicted in a
particular partial wave by conventional $qqq$ models. Doing so will
require careful multi-channel analysis of reactions involving many
different initial and final states~\cite{Dytman}. Another possibility
is that such states will have characteristic electromagnetic
production amplitudes~\cite{Roper}. If hybrid baryons obey similar
decay selection rules to hybrid mesons~\cite{pageselectionrule}, they
may be distinguishable based on their strong decays. This work
concentrates on a determination of their masses and quantum numbers, in
an approach where the physics of the confining interaction defines the
relevant gluonic degrees of freedom.

Hybrid baryons have been examined using QCD sum rules~\cite{QCDsr} and
in the large number of colors (large $N_c$) limit of
QCD~\cite{yan}.  One approach to modeling the structure of hybrid
baryons (not taken here) is to view them as bound states of three
quarks and a `constituent' gluon. Hybrid baryons have been constructed
in the MIT bag model~\cite{bag} by combining a constituent gluon in
the lowest energy transverse electric mode with three quarks in a
color-octet state, to form a color-singlet state. With the assumption
that the quarks are in an $S\,$-wave spatial ground state, and
considering the mixed exchange symmetry of the octet color wavefunctions
of the quarks, bag-model constructions show that adding a $J^P=1^+$
gluon to three light quarks with total quark spin 1/2 yields both $N$
($I=\frac{1}{2}$) and $\Delta$ ($I=\frac{3}{2}$) hybrids with
$J^P=\sfrac{1}{2}^+$ and $\sfrac{3}{2}^+$. Quark spin 3/2 hybrids are
$N$ states with $J^P=\sfrac{1}{2}^+$, $\sfrac{3}{2}^+$, and
$\sfrac{5}{2}^+$. Energies are estimated using the usual bag
Hamiltonian plus gluon kinetic energy, additional color-Coulomb
energy, and one-gluon exchange plus gluon-Compton O($\alpha_s$)
corrections. Mixings between $q^3$ and $q^3g$ states from gluon
radiation are evaluated. If the gluon self energy is included, the
lightest $N$ hybrid state has $J^P=\frac{1}{2}^+$ and a mass between
that of the Roper resonance and the next observed $J^P=\sfrac{1}{2}^+$
state, $N(1710)$. A second $J^P=\frac{1}{2}^+$ $N$ hybrid and a
$J^P=\frac{3}{2}^+$ $N$ hybrid are expected to be 250 MeV heavier,
with the $\Delta$ hybrid states heavier still. A similar mass estimate
of about 1500 MeV for the lightest hybrid is attained in the QCD sum
rules calculation of Ref.~\cite{QCDsr}.

For this reason there has been considerable interest in the presence
or absence of light hybrid states in the $P_{11}$ and other
positive-parity partial waves in $\pi N$ scattering.  Interestingly,
quark potential models which assume a $q^3$ structure for the Roper
resonance~\cite{seeCI,capstick86} predict an energy which is roughly
100 MeV too high, and the same is true of the $\Delta(1600)$, the
lightest radial recurrence of the ground state $J^P=\frac{3}{2}^+$
$\Delta(1232)$. Furthermore, models of the electromagnetic couplings
of baryons have difficulty accommodating the substantial Roper
resonance photocoupling extracted from pion photoproduction
data~\cite{RoperPC}. Evidence for two resonances near 1440 MeV in the
$P_{11}$ partial wave in $\pi N$ scattering was cited~\cite{twopoles},
which would indicate the presence of more states in this energy region
than required by the $q^3$ model, but this has been interpreted as due
to complications in the structure of the $P_{11}$ partial wave in this
region, and not an additional $qqq$
excitation~\cite{CutkoskyP11}. Recent calculations~\cite{Krewald} of
$N\pi \to N\pi\pi$ reaction observables incorporating baryon-meson
dynamics are able to describe this reaction in the Roper resonance
region in the absence of a $qqq$ excitation, and find a
dynamically-generated pole at the mass of the Roper resonance. Given
this complicated structure, it is perhaps not surprising that there
are difficulties in describing the photocouplings of this state within
a simple three-quark picture.

\begin{figure}
\includegraphics[width=6cm,angle=0]{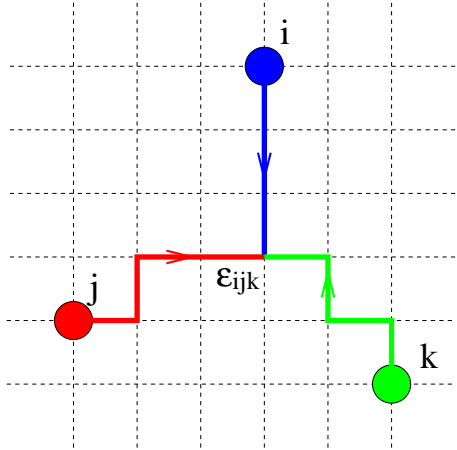}
\caption{\label{lattice_bar} A possible configuration of flux lines in a
baryon on the lattice.}
\end{figure}

The motivation of this work is to build a model consistent with
predictions from QCD lattice gauge theory, based on the Isgur-Paton
non-relativistic flux-tube model~\cite{paton85}. This model is
motivated from the strong coupling limit of the Hamiltonian lattice
gauge theory formulation of QCD (HLGT). This strong coupling limit
predicts linear confinement in mesons proportional to the expectation
of the Casimir operator for color charges, which has been verified in
lattice QCD~\cite{deldar}. In conventional baryons, in the limit of
heavy quarks, the static confining potential has been shown in lattice
calculations~\cite{bali} to be consistent with that given by a
minimum-length configuration of flux tubes meeting in a Y-shaped
configuration at a junction (see Fig.~\ref{lattice_bar}), and {\it
not} consistent with two-body confinement, where a triangle of tubes
would connect the quarks in a $\Delta$ configuration~\cite{Simonov}. It
is possible to experimentally examine this configuration in studies of
baryon production in the central rapidity region of ultrarelativistic
nucleon and nuclear collisions \cite{kharseev}.

This structure of the glue, where the gluon degrees of freedom
condense into flux tubes, is very different from the constituent-gluon
picture of the bag model and large-$N_c$ constructions. Substantial
progress has been made in recent years in understanding conventional
baryons by studying the large $N_c$ limit of QCD. However, the large
$N_c$ limit does not necessarily provide model-independent results on
hybrid baryons.  Hybrid baryons in the large $N_c$ limit consist of a
single gluon and $N_c$ quarks~\cite{yan}. Even in the case of physical
interest, $N_c=3$, this does not correspond to the description of
hybrid baryons presented here, where is is argued that the dynamics
relevant to the structure of hybrids are that of confinement, where
numerous gluons have collectively condensed into flux tubes. Since the
color structure of a hybrid baryon determines (through the Pauli
principle) its flavor structure, the $N_c=3$ case gives rise to states
with different quantum numbers than in our approach. It has been shown
by Swanson and Szczepaniak~\cite{swanson98} that a constituent-gluon
model is not able to reproduce lattice QCD data~\cite{morningstar97bag}
for hybrid-meson potentials at large interquark seperations. In
addition, the flux-tube model hybrid-meson potential is consistent at
large interquark separations with that evaluated from lattice
QCD~\cite{paton85,adia}.

Hybrid baryons are constructed here in the adiabatic approximation,
where the quarks do not move in response to the motion of the glue,
apart from moving with fixed interquark distances in order to maintain
the center-of-mass position. The effect of the motion of the glue in
hybrid baryons (and the zero-point motion of the glue in conventional
baryons) is to generate a confining potential in which the quarks are
allowed to move. This differs from that found from multiplying the sum
of the lengths of the strings (``triads'') connecting the quarks to
the junction by the string tension. The adiabatic approximation is
exact only in the heavy quark limit, although the success of quark
model phenomenology of conventional mesons and baryons implies that
there is a close relation between heavy-quark and light-quark
physics. A modified adiabatic approximation is employed, which can be
shown to give exact energies and wave functions for specific dynamics
even for light quarks~\cite{page00}. Moreover, a modified adiabatic
approximation has been shown to be good for properties of light quark
mesons in the flux-tube model~\cite{merlin85}.

\begin{figure}
\includegraphics[width=15cm,angle=0]{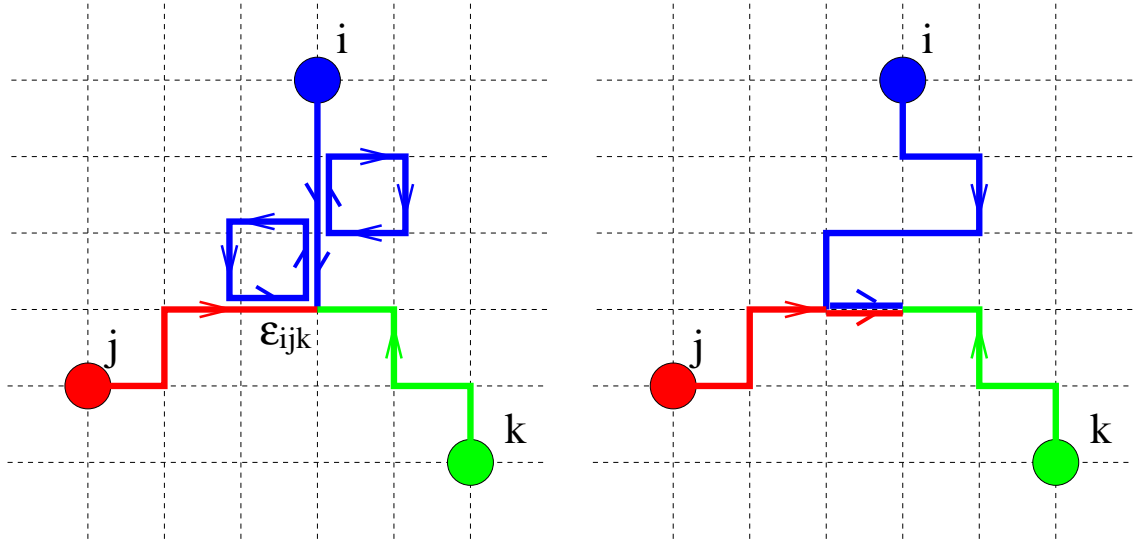}
\caption{\label{lattice_plaq} Flux lines in a baryon on the lattice,
illustrating the application of a pair of plaquette operators, the
upper operator moving one flux line perpendicular to its starting
position, and the lower attempting to move the junction.}
\end{figure}

The model is motivated from the strong coupling limit of HLGT, where
there are ``flux lines'' which play the role of glue.  In the
adiabatic approximation operators which make the quarks move are
neglected.  The plaquette operator corrects the strong coupling limit,
and induces motion of the ``flux lines'' between the quarks and the
junction perpendicular to their rest positions. The flux lines are
modeled by equally spaced ``beads'' of identical mass, so that the
energy of each flux line is proportional to the number of beads, and
hence its length. The spacing of the beads along the rest positions of
the flux lines can be thought of as a finite lattice spacing, and the
beads are allowed to move perpendicular to their rest position. The
beads are attracted to each other by a linear potential, and the
resulting discretized flux lines vibrate in various modes.  Global
color invariance requires that the three flux lines emanating from the
quarks meet at a junction, which is also modeled by a bead. It is
shown in HLGT that a single plaquette operator can move the junction
and retain the Y-shaped string with the links in their ground state
(see Fig.~\ref{lattice_plaq} and Appendix~\ref{junct}), so that the
junction may have a similar mass to the beads. However, for generality
we allow the junction to have a different mass associated with it
than that of the other beads.

The final picture of both conventional and hybrid baryons is that of
three quarks, connected via a line of beads to the junction in a
Y-shaped configuration. The potential between neighboring beads is
linear. The adiabatic approximation is used, so that the string is
assumed to adjust its state quickly in response to motion of the
quarks, thus generating a potential in which the quarks move, in both
conventional and hybrid baryons. The motion of the quarks in these
potentials is then solved for variationally. A brief outline of this
approach is given in Ref.~\cite{CP1}. The purpose of the present paper
is to describe the model and the calculation of hybrid baryon masses
in much more detail, and to put this work in the context of recent
advances in lattice gauge theory.

\begin{table}
\caption{\small Notations frequently used in the main text.} 
\label{tab1}
\begin{tabular}{ll}
\hline
$b$ & string tension\\[-6 pt]
$\thetaj$ & angle between triads suspended at the equilibrium junction 
position \\[-6 pt]
$i = 1,2,3$  & quark or triad label\\[-6 pt]
$N_i$  &  number of beads on triad $i$\\[-6 pt]
$n = 1,N_i$  & counts the beads on triad $i$ from  quark $i$ to the 
junction\\[-6 pt]
$m = 1,N_i$  & counts the modes of the triad $i$\\[-6 pt]
$M_i$ &   mass of quark $i$\\[-6 pt]
$m_b$ &   mass of the beads\\[-6 pt]
$m_J$  &  mass of the junction\\[-6 pt]
$l_i$ &   distance from the equilibrium junction position to quark $i$\\[-6 pt]
$\hat{\bl}_i$ &   direction from the equilibrium junction position to 
quark $i$\\[-6 pt]
$\br_i$  &    position of quark $i$\\[-6 pt]
$q_{\parallel m}^{i}$ & amplitude of mode $m$ on triad $i$ in the $QQQ$ 
plane, but perpendicular to $\hat{\bl}_i$\\ [-6 pt]
$q_{\perp m}^{i}$ &  amplitude of mode $m$ on triad $i$
perpendicular to the $QQQ$ plane\\ [-6 pt]
${\bf r} = (x,y,z)$  &  cartesian coordinates of the junction\\[-6 pt]
$\etahat_{\pm},\; \etahat_z$ & junction oscillation directions parallel 
and perpendicular to the $QQQ$ plane, respectively\\[-6 pt]
$\omega_{\pm},\; \omega_z$ & junction oscillation frequencies parallel 
and perpendicular to the $QQQ$ plane, respectively\\
\hline
\end{tabular}
\end{table}
The rest of the paper is organized as
follows. Section~\ref{fluxdynamics} describes the dynamics of the
flux tubes in various quark configurations, with analytic solutions in
special cases. Section~\ref{fluxsymmetry} discusses the quark-label
exchange symmetry, parity, and chirality of the flux configuration. In
Section~\ref{qmnos}, the orbital angular momentum and color of the
flux, and the combined quark and flux wave function are
constructed. The following Section~\ref{fluxpot} describes the
potential in which the quarks move, which includes the energy of the
flux. Numerical estimates of the masses of hybrid baryons are given in
Section~\ref{numerical}. In Sections~\ref{discuss} and~\ref{conc} further
discussions and conclusions are given.

\section{Flux Dynamics\label{fluxdynamics}}

\begin{figure}
\includegraphics[width=8cm,angle=0]{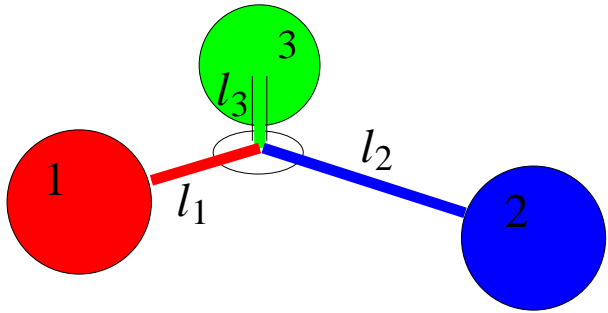}
\caption{\label{bar_conf} Flux configuration when none of the angles
in the triangle joining the quarks are larger than 120$^{\rm o}$.}
\end{figure}

Denote by $\theta_{132}$ the angle between the line from quark 1 to 3
and that from quark 2 to 3. If $\theta_{123}$, $\theta_{132}$, and
$\theta_{213}$ are all smaller than $120^{\rm o}$, the flux is in its
equilibrium (lowest-energy) configuration when the junction is located
such that there are angles of $120^{\rm o}$ between each of the triads
that connect each of the quarks to the junction, and the beads all lie
on the triads.  In this lowest energy configuration, the string lies
in the plane defined by the three quarks, denoted by the $QQQ$
plane. The angle between the line from any two quarks to the junction
equilibrium position $\req$ is $120^{\rm o}$, which is denoted by
$\thetaj=120^{\rm o}$ (see Fig.~\ref{bar_conf} and Table~\ref{tab1}).
%
%
In terms of the lengths $l_i$ of the lines from the $i$-th quark to
the junction, and the quark positions ${\bf r}_i$, the equilibrium
junction position is
\beqn 
\label{req}
\req = \frac{l_2l_3\br_1+ l_1l_3\br_2+ l_1l_2\br_3}{l_2l_3+
l_1l_3+ l_1l_2}.
\eeqn
The position vectors of the quarks relative to the junction
equilibrium position are $\bl_i = \br_i - \req$.

\begin{figure}
\includegraphics[width=8cm,angle=0]{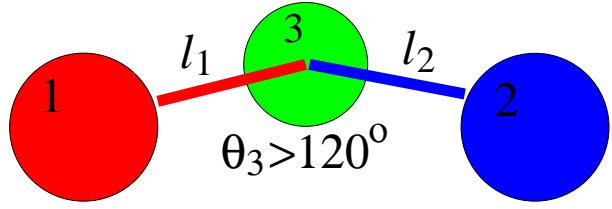}
\caption{\label{bar_hybrid_120} Flux configuration when one of the
angles in the triangle joining the quarks, here $\theta_{132}$, is
larger than 120$^{\rm o}$.}
\end{figure}

If one of $\theta_{123}$, $\theta_{132}$, and $\theta_{213}$ is larger
than or equal to $120^{\rm o}$, the equilibrium configuration of the
flux is not this Y-shaped configuration. If $\theta_{ijk}>120^{\rm
o}$, the lowest energy configuration has the junction at the position
of quark $j$ (see Fig.~\ref{bar_hybrid_120}). This situation is
denoted by $\thetaj>120^o$. In what follows the case where
$\theta_{132} > 120^{\rm o}$ is analyzed, but the formulae for the
other cases follow by the appropriate label exchange.

\begin{figure}
\includegraphics[width=12cm,angle=0]{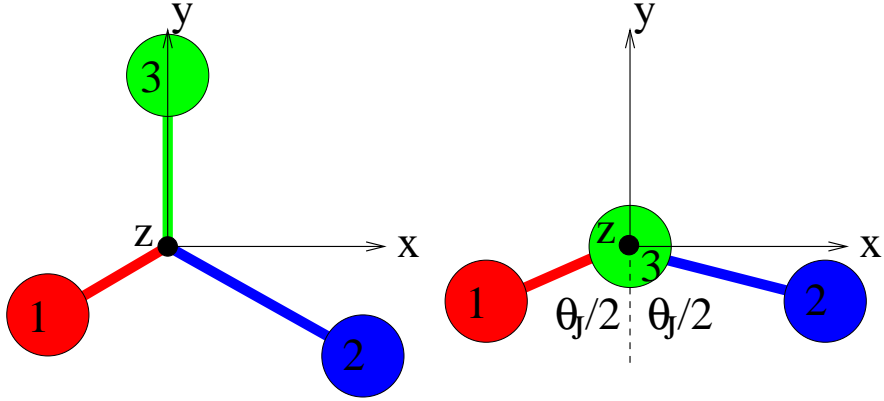}
\caption{\label{hyb_coords} Coordinate system used to describe motion
of the flux in the QQQ plane, for configurations with (a) all of
$\theta_{123}$, $\theta_{132}$, and $\theta_{213}$ less than 120$^{\rm
o}$ and (b) with $\theta_J=\theta_{132}$ larger than 120$^{\rm o}$.}
\end{figure}

An axis system is chosen as indicated in Figure~\ref{hyb_coords}.
%
%
This defines normalized $\hat{\bf x}$ and $\hat{\bf y}$, which can
also be written in the $\thetaj=120^{\rm o}$ case as
\beqn
\label{xhat}
\hat{\bf x} = -\frac{\hat{\bl}_1-\hat{\bl}_2}{\sqrt{3}}
\hspace{1cm} \hat{\bf y} =
-\frac{\hat{\bl}_1+\hat{\bl}_2-2\hat{\bl}_3}{3},
\eeqn
where the $\hat{\bl}_i$ are unit vectors along the triads, and
$\hat{\bl}_1+\hat{\bl}_2+\hat{\bl}_3 = 0$, so that $\hat{\bf y}$
equals $\hat{\bl}_3$ and $\hat{\bf x}$ is perpendicular to
$\hat{\bl}_3$.  The third triad lies on the positive $y$-axis and the
other two triads are $120^{\rm o}$ on either side of the $y$-axis:
triad one on the left-hand and triad two on the right-hand side.  It
is assumed that the sum of the masses of the beads and the junction is
the energy of the flux configuration in its equilibrium position, so
that
\beqn 
\label{masssum} 
m_b \sum_{i=1}^3 N_i + m_J = b\sum_{i=1}^3 l_i,
\eeqn
which implies
\beqn
m_b=\frac{b\sum_{i=1}^3
l_i}{\sum_{i=1}^3 N_i + \frac{m_J}{m_b}},
\eeqn
where $N_i$ is the number of beads on triad $i$.  Note from the
above that the bead mass $m_b$ is determined by the string tension
$b$, the triad lengths $l_i$, and the ratio $m_J/m_b$, and so should
not be regarded as an independent parameter.

\subsection{Hamiltonian for $\thetaj=120^o$ case}

\begin{figure}
\includegraphics[width=8cm,angle=0]{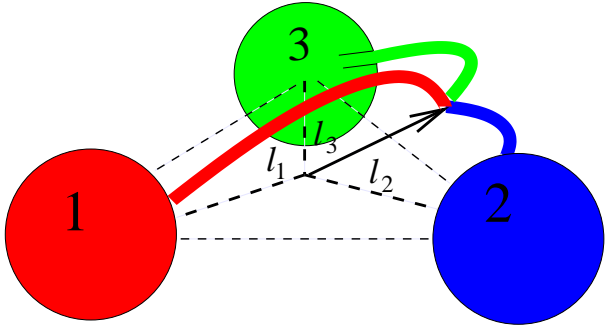}
\caption{\label{baryon_hybrid} Motion of the flux configuration
decomposed into ``junction motion'' and ``bead motion''.}
\end{figure}

\begin{figure}
\includegraphics[width=8cm,angle=0]{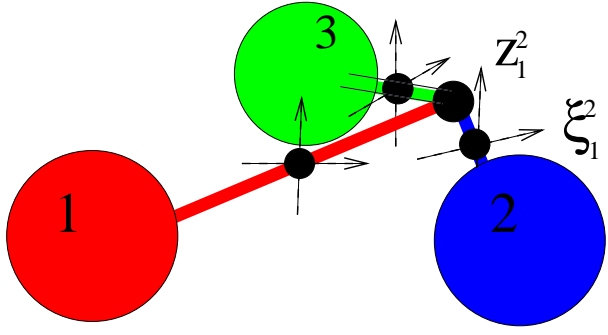}
\caption{\label{bar_hybrid_beads} Coordinates used to describe the
motion of the beads in the presence of a displaced junction.}
\end{figure}

The flux configuration is made dynamical by allowing the junction and
the beads to vibrate with respect to their equilibrium configuration.
There are two important motions which are expected to have physical
significance: (1) the motion of the junction perpendicular to and
within the plane relative to its rest position, denoted ``junction
motion'', and (2) the motion of the beads in the two directions
perpendicular to the line connecting the quark to the junction, called
``bead motion'', as illustrated in Fig.~\ref{baryon_hybrid}. 
%
%
The bead motion coordinates are not their positions, but the
oscillating-wave amplitudes (defined in Appendix~\ref{app1}) of the
beads relative to their rest positions on the triads.  It is important
for what follows that the bead position coordinates are defined
relative to their rest positions on the triads between the quarks and
junction, which have followed the junction motion (see
Fig.~\ref{bar_hybrid_beads}).
%
%
The Hamiltonian is written in terms of the junction and bead motion
coordinates. In what follows the small-oscillations approximation is
used, where the beads and junction remain close to their positions in
the equilibrium configuration. This approximation is used to motivate
the basis of the subsequent numerical treatment, which is that it is a
reasonable approximation to treat the flux motion as that of the
junction, with an effective mass which depends on the equilibrium
lengths of the triads, among other quantities. In the numerical
treatment that follows, the restriction to small oscillations is
removed.

As the string moves, the quarks are allowed to move with fixed
positions relative to each other, in order to keep the center of mass
fixed. This is called the ``redefined adiabatic'' approximation. By
working in this approximation, some but not all non-adiabatic effects
are incorporated.

The flux Hamiltonian for the $\thetaj=120^{\rm o}$ case in the
redefined adiabatic approximation is (see Appendix \ref{app1})
\beqna \label{ham1}
H_{\mbox{\small flux}}&=&\frac{1}{2}\meff{\bf \dot{r}}^2
+\vbead^{\mbox{\small sm.osc.}}
\nonumber\\
&+&\frac{1}{2}\beff \sum_{i=1}^3 \frac{N_i+1}{2} \sum_{m=1}^{N_i} 
\left[
({{\dot{q}}_{\parallel m}^{i})}^2+{({\dot{q}}_{\perp m}^{i})}^2
\right]+\frac{b}{2}\sum_{i=1}^3 l_i \sum_{m=1}^{N_i} 
\frac{1}{2}
(\omega^i_m)^2\left[(q_{\parallel m}^{i})^2+(q_{\perp m}^{i})^2\right]
\nonumber \\
&+&m_b\; {\bf \dot{r}} \cdot \sum_{i=1}^{3}\sum_{m=1}^{N_i} 
\left[\beta^i_m
- \frac{m_b}{2}\;\frac{\sum_{k=1}^3N_k+2\frac{m_J}{m_b}}
{\sum_{k=1}^3(bl_k+M_k)} \;
\alpha^i_m\right]
\left(e^i_x{\dot{q}}_{\parallel
m}^{i},e^i_y{\dot{q}}_{\parallel m}^{i},{\dot{q}}_{\perp
m}^{i}\right)
\nonumber  \\
&-&\frac{m_b^2}{2}\frac{1}{\sum_{i=1}^3(bl_i+M_i)} 
\left\{ 
\sum_{i=1}^3  \sum_{n\neq m}^{N_i} \alpha^i_n\alpha^i_m 
({\dot{q}}_{\parallel n}^i {\dot{q}}_{\parallel m}^i +
{\dot{q}}_{\perp n}^i {\dot{q}}_{\perp m}^i)
\right.
\nonumber  \\ 
& &
+ \sum_{i\neq j}^3 \sum_{n=1}^{N_i}\sum_{m=1}^{N_j}\alpha^i_n\alpha^j_m 
\left[\lpmb{e}^i\cdot\lpmb{e}^j{\dot{q}}_{\parallel n}^i
{\dot{q}}_{\parallel m}^j + {\dot{q}}_{\perp n}^i 
{\dot{q}}_{\perp m}^j\right]
\nonumber \\
& &\left.
+\sum_{i=1}^3\sum_{m=1}^{N_i}\left[(\alpha^i_m)^2-\frac{N_i+1}{2}
\right]
\left[({\dot{q}}_{\parallel m}^i)^2+({\dot{q}}_{\perp m}^i)^2\right].
\right\}
\eeqna
The first term in Eq.~\ref{ham1} is the kinetic energy of the
junction, with an effective mass of 
\beqn\label{meff}
\meff\equiv m_b \left[ \sum_{i=1}^{3}\frac{N_i(2 N_i+1)}{6(
N_i+1)}+\frac{m_J}{m_b}-\frac{m_b{(\sum_{i=1}^3N_i+2\frac{m_J}{m_b})}^2}{4(
\sum_{i=1}^3(bl_i+M_i))} \right],
\eeqn
where the last term in Eq.~\ref{meff} arises from the center-of-mass
correction, and the first from the trivial motion of the beads which
accompanies motion of the junction. The second term in Eq.~\ref{ham1}
is the potential energy of the junction in the small-oscillations
approximation, given in terms of the coordinates ${\bf
r}\equiv(x,y,z)$ defined in Fig.~\ref{hyb_coords} by
\beqna
\label{vbead3}
\vbead^{\mbox{\small sm.osc.}} & \equiv &
b\sum_{i=1}^3 l_i \nonumber\\
&\hspace{-4cm} +&\hspace{-2cm} \frac{b}{2} \left[
x^2\left(\frac{1}{4l_1}+\frac{1}{4l_2}+\frac{1}{l_3}\right)
+\frac{3}{4}y^2\left(\frac{1}{l_1}+\frac{1}{l_2}\right)
-\frac{\sqrt{3}}{2}xy
\left(\frac{1}{l_1}-\frac{1}{l_2}\right)+z^2\left(\frac{1}{l_1}+\frac{1}{l_2}
+\frac{1}{l_3}\right)
\right].
\eeqna
The third and fourth terms in Eq.~\ref{ham1} are the kinetic and
potential energies of the beads, respectively, written in terms of the
effective mass of the beads, including a center-of-mass correction,
\beqn
\label{beff}
\beff\equiv m_b \left (1- \frac{m_b}{\sum_{i=1}^3(bl_i+M_i)}\right),
\eeqn
and the frequency of the $m$-th normal mode on the $i$-th triad,
\beqn
\omega^i_m \equiv  \frac{2(N_i+1)}{l_i}\sin \frac{m\pi}{2(N_i+1)}.
\eeqn

The fifth and sixth terms in Eq.~\ref{ham1} represent interactions
between the junction and the beads, and interactions among the
beads, respectively. In these terms the vectors $\lpmb{e}^i$ are
defined to be perpendicular to $\hat{\bl}_i$, so that
\beqn \label{ees}
\lpmb{e}^1=\left(-\frac{1}{2},\frac{\sqrt{3}}{2},0\right), 
\hspace{.7cm}
\lpmb{e}^2=\left(-\frac{1}{2},-\frac{\sqrt{3}}{2},0\right),
\hspace{.7cm}
\lpmb{e}^3=\left(1,0,0\right),
\hspace{.7cm} \lpmb{e}^i \cdot \lpmb{e}^j = -\frac{1}{2}.
\eeqn
In terms of the amplitudes $\lpmb{q}_m^i$ of the $m$-th normal mode on
the $i$-th triad, where displacements of the beads along the triads
are not allowed, the coordinates $q_{\parallel m}^{i}=\lpmb{e}^i\cdot
\lpmb{q}_m^i$ are the projections in the QQQ plane, and the
coordinates $q_{\perp m}^{i}$ are the projections of these amplitudes
out of the QQQ plane. Finally, the $\beta^i_m$ and $\alpha^i_m$ are
the sums
\beqn
\beta^i_m \equiv  \sum_{p=1}^{N_i}\frac{p}{N_i+1}\sin\frac{mp\pi}{N_i+1}
= \frac{(-1)^{m+1}}{2}\cot\frac{m\pi}{2(N_i+1)}
\eeqn
and
\beqn
\alpha^i_m \equiv  \sum_{p=1}^{N_i}\sin\frac{mp\pi}{N_i+1}
= \begin{cases}
0 & m= {\rm even}\\
\cot\frac{m\pi}{2(N_i+1)} & m= {\rm odd}.
\end{cases}
\eeqn

The above demonstrates that the Hamiltonian can be seperated into
three parts. The first (the first two terms in Eq. \ref{ham1})
corresponds to the motion of the junction in the absence of beads,
with an effective junction mass related to its own mass and the bead
mass, with a center-of-mass correction for finite quark masses. The
second part (terms three and four) is the independent motion of the
beads on the three triads with respect to a fixed junction, with a
bead mass also corrected for center-of-mass motion for finite quark
masses.  There is also an ``interaction'' part where the junction
interacts with the various bead modes (term five), where the bead
modes associated with the same quark interact with each other, and
where the modes on triads corresponding to different quarks interact
with each other (term six). Note that these bead self-interactions
(term six) vanish for infinite quark masses.

Because the quarks move with fixed relative positions only to maintain
the center of mass position in the presence of a moving junction and
beads, there are no quark kinetic terms in this string Hamiltonian, and
there is no sense in which the quarks acquire mass from the beads,
i.e. constituent quark masses are not derived from current quark
masses.

Note that the model predicts that for bead motion in the
small-oscillations approximation the potential has the customary, and
phenomenologically important~\cite{capstick86}, linear potential term
$b\sum_{i=1}^3 l_i$ (see Eq. \ref{vbead3}).  For large $l_i$, where
the small-oscillations approximation becomes exact, the potential is
just the linear term expected in any string model.  This potential is
a prediction of the model, not an ansatz.  In the numerical work that
follows, the small-oscillations approximation for the junction will be
removed to yield the potential energy in the absence of beads, i.e.
\beqna \label{vbead2} \lefteqn{\vbead\equiv b\left (
\sqrt{\left(\frac{\sqrt{3}}{2}l_1+x\right)^2
+\left(\frac{l_1}{2}+y\right)^2+z^2} 
\right. \nonumber } \\ & & \left.
+ \sqrt{\left(\frac{\sqrt{3}}{2}l_2-x\right)^2
+\left(\frac{l_2}{2}+y\right)^2+z^2}
+ \sqrt{x^2+(l_3-y)^2+z^2} \right ).
\eeqna

\subsection{Hamiltonian for $\thetaj>120^o$ case \plabel{gsec}}

The correct expression in the small-oscillations approximation for the
flux Hamiltonian for the case $\thetaj>120^o$ can be obtained (see
Appendix \ref{app1}) by setting $l_3 = N_3 = 0$ in $H_{\mbox{\small
flux}}$, $\meff$, and $\beff$ in Eqs.~\ref{ham1},~\ref{meff},
and~\ref{beff}, and $q_{\parallel m}^{3}$, $q_{\perp m}^{3}=0$ in
$H_{\mbox{\small flux}}$ in Eq.~\ref{ham1}, and changing the junction
potential energy in Eq.~\ref{vbead3} to
\beqna
\plabel{V120}
\lefteqn{\vbead^{\mbox{\small sm.osc.}} \equiv b\sum_{i=1}^2 l_i \;
+ 2 b y \cos\frac{\thetaj}{2}  + b\sqrt{x^2+y^2+z^2} \nonumber } \\
 & &  +\; \frac{b}{2} \left( \left[x^2\cos^2\frac{\thetaj}{2} 
+y^2\sin^2\frac{\thetaj}{2}+z^2\right]\left\{\frac{1}{l_1}
+\frac{1}{l_2}\right\} 
- 2  xy\sin\frac{\thetaj}{2}\cos\frac{\thetaj}{2}
\left\{\frac{1}{l_1}-\frac{1}{l_2}\right\}\right).
\eeqna
Note that the third term above, which is the length of the third triad
when the junction has moved, cannot be expanded in the
small-oscillations approximation. The vectors $\lpmb{e}^i$ become
\beqn 
\plabel{eet}
\lpmb{e}^1=\left(-\cos\frac{\thetaj}{2},\sin\frac{\thetaj}{2},0\right),
 \hspace{.7cm}    
\lpmb{e}^2=\left(-\cos\frac{\thetaj}{2},-\sin\frac{\thetaj}{2},0\right).
\eeqn
In the numerical work that follows, the small-oscillations
approximation for the junction will be removed to yield the potential
energy in the absence of beads,
\beqna 
\plabel{vbead1} 
\lefteqn{\vbead\equiv b\left (
\sqrt{(l_1\sin\frac{\thetaj}{2}+x)^2+(l_1\cos\frac{\thetaj}{2}+y)^2+z^2}
\right. \nonumber } \\ & & \left.
+ \sqrt{(l_2\sin\frac{\thetaj}{2}-x)^2+(l_2\cos\frac{\thetaj}{2}+y)^2+z^2}
+ \sqrt{x^2+y^2+z^2} \right ).
\eeqna

\subsection{Approximate flux Hamiltonian}

It will now be demonstrated that the interaction terms in
Eq. \ref{ham1} (terms five and six) give a minor contribution in the
small-oscillations approximation.  The free parameters in the model
(and the values initially used for the numerical simulation) are the
string tension ($0.18$ GeV$^2$), the ratio of the junction and bead
masses (1), and the quark masses ($0.33$ GeV). The simulation is
performed with one bead between each quark and the junction, and the
quarks at first form an equilateral triangle with the lengths of the
triads given a typical value of $2.5$ GeV$^{-1}$.

For the purposes of this demonstration, the problem is first solved
numerically without approximations, by solving the classical
Euler-Lagrange equations of motion rather than using quantum
mechanics. This solution should provide a good indicator of how the
mode frequences with and without the interaction Hamiltonian
compare. The mode frequencies parallel and perpendicular to the plane
are (in GeV)
\begin{tabbing} \label{bnl2}
XXXXXXXXXXXXXXXX\=XXXXXX\=XXXXXX\=XXXXXX\=XXXXXX\=XXXXXX\kill 
Parallel\>{\bf 0.607}\>{\bf 0.607}\>0.924\>1.08\>1.08\\
Perpendicular\>{\bf 0.828}\>\>0.924\>0.924\>1.37
\end{tabbing}
where the lowest frequencies have been identified in bold face. If
the interaction Hamiltonian is set to zero, the number of modes does not
change, as the number of degrees of freedom (three junction
coordinates and two transverse bead coordinates per bead) is
unchanged. Since the junction and bead degrees of freedom become
uncoupled, it is possible to identify modes involving junction motion
and those involving bead motion. The mode frequencies (in GeV)
corresponding to the junction motion (bold) and bead vibrations
are
\begin{tabbing} \label{bnl2}
XXXXXXXXXXXXXXXX\=XXXXXX\=XXXXXX\=XXXXXX\=XXXXXX\=XXXXXX\kill 
Parallel\>{\bf 0.614}\>{\bf 0.614}\> 1.00\> 1.00\> 1.00\\
Perpendicular\>{\bf 0.869}\> \> 1.00\> 1.00\> 1.00
\end{tabbing}
The similarity of the frequencies of the lowest energy modes in this
approximation to those arising from the full Hamiltonian (a deviation
of 1\% in the case of the modes with motion parallel to the QQQ plane,
and 5\% for the mode perpendicular to the plane) shows that for these
lowest energy modes the interaction Hamiltonian can be safely
neglected. In retrospect, the reason for this is because of the choice
of physically appropriate coordinates for the problem, i.e. the
junction coordinates and the coordinates of the beads transverse
to the triads joining the junction to the quarks (see
Fig.~\ref{bar_hybrid_beads}).

To ensure that this result is not dependent on this choice of QQQ
configuration or the parameters in the Hamiltonian, the parameters
were varied independently around the central values used above. Quark
masses up to the charm quark mass of $1.5$ GeV were used, the ratio of
the junction to the bead mass was taken up to 10, and the triads were
given lengths from $0.5 - 5$ GeV$^{-1}$, and cases with unequal
lengths were tested. The percentage difference between all nine mode
frequencies calculated with the full Hamiltonian and with the
interaction terms neglected is shown for selected parameters in
Table~\ref{tabb}. The largest error for the two lightest parallel
modes and the lightest perpendicular mode, shown in bold face, is
7\%. This demonstrates that, to a good approximation, the dynamics of
the three lowest frequency modes can be simplified to junction and
bead motion which are independent of one another. The bead motion on
various triads, and bead motion in various modes on the same
triad are to a good approximation independent of each other.
\begin{table}[t]
\begin{center}
\caption{\small Percentage differences between all nine mode
frequencies obtained in the one bead per triad problem for the full
Hamiltonian (Eq.~\ref{ham1}) and the Hamiltonian with the
interaction terms (the last two terms in Eq.~\ref{ham1})
neglected, for selected parameters. The first group of five percentage
differences is for parallel mode frequencies, the last four for
perpendicular mode frequencies. Within each group the percentage
differences are displayed from left to right for in ascending order of
mode frequency. Quark masses are in GeV and triad lengths in
GeV$^{-1}$.}
\vskip 0.5 cm
\plabel{tabb}
\begin{tabular}{|l|l|l|r|l|l|l||r|r|r|r|r||r|r|r|r|}
\hline 
$M_1$ & $M_2$ & $M_3$ & $\frac{m_J}{m_b}$ & $l_1$ & $l_2$ & $l_3$ & \multicolumn{9}{c|}{Percentage difference (\%)} \\
\hline \hline 
0.33 & 0.33 & 0.33 & 1 & 2.5 & 2.5 & 2.5 & \bf 1 & \bf 1 & 8 & 8 & 8 & \bf 5 & 8 & 8 & 37 \\
1.5  & 0.33 & 0.33 & 1 & 2.5 & 2.5 & 2.5 & \bf 2 & \bf 2 & 5 &11 &11 & \bf 7 & 5 & 5 & 40 \\
0.33 & 0.33 & 0.33 & 10& 2.5 & 2.5 & 2.5 & \bf 0 & \bf 0 & 2 & 1 & 1 & \bf 0 & 2 & 2 &  5 \\
0.33 & 0.33 & 0.33 & 1 & 0.5 & 2.5 & 2.5 & \bf 1 & \bf 6 & 9 & 2 & 4 & \bf 7 & 0 &10 & 11 \\
0.33 & 0.33 & 0.33 & 1 &   5 & 2.5 & 2.5 & \bf 1 & \bf 1 & 0 & 2 & 7 & \bf 3 & 2 & 8 & 28 \\
0.33 & 0.33 & 0.33 & 1 & 0.5 & 2.5 &   5 & \bf 1 & \bf 1 & 2 & 1 & 3 & \bf 2 & 0 & 4 & 10 \\
\hline 
\end{tabular}
\end{center}
\end{table}

The frequencies can be followed from the non-interacting case as
interactions are turned on, and level crossing does not occur. Hence
mode frequencies for the fully-interacting Hamiltonian can be uniquely
associated with modes frequencies obtained with interactions
neglected. The lowest frequency is {\it always} associated with the
lowest junction excitation. However, the next lowest frequency can be
associated with the second junction excitation or with a bead
excitation along a triad when the QQQ configuration is asymmetric.
This work focuses on the lowest lying excitation of the flux
configuration, always corresponding to junction motion, but it should
be kept in mind that the next lowest hybrid baryon may involve bead
excitation.

The equal-mass three-bead model is unrealistic when one of the triads
is short, since the mass of the bead on the short triad is not
representative of the energy stored in the triad on which it lies.
An alternative model has been considered where the bead mass is taken
to be proportional to the length of the triad, but the sum of the bead
masses and junction mass still equals the energy stored in the
Y-shaped string configuration (Eq.~\ref{masssum} with $N_i=1$). In
this model it is found that the low-lying frequencies for the full
Hamiltonian are very similar to the former model, with similar small
errors induced by neglecting the interaction terms in the Hamiltonian
of Eq.~\ref{ham1}.

\subsection{Analytical solution of the flux Hamiltonian}

For the Hamiltonian in Eq.~\ref{ham1} (the $\thetaj= 120^o$ case) the
last two terms have been shown to be negligible for the lowest
frequency in the small-oscillations approximation, in the case where
there is one bead on each triad, and these terms are neglected in what
follows. What follows is, therefore, based on the approximate
$\thetaj= 120^o$ Hamiltonian
\beqna 
\plabel{ham2}
{\tilde H}_{\mbox{\small flux}}
&\equiv&
\frac{1}{2}\meff{\bf \dot{r}}^2+\vbead
+\frac{1}{2}\beff \sum_{i=1}^3\frac{N_i+1}{2}
\sum_{m=1}^{N_i} 
\left[(\dot{q}_{\parallel m}^i)^2+(\dot{q}_{\perp m}^i)^2\right]
 \nonumber \\ 
& &+\frac{b}{2}\sum_{i=1}^3l_i\sum_{m=1}^{N_i} 
\frac{1}{2}(\omega^i_m)^2\left[(q_{\parallel
m}^{i})^2+(q_{\perp m}^{i})^2\right],
\eeqna
with $\vbead$ given by Eq.~\ref{vbead2}. Note that since the $q_{\parallel
m}^{i}$, $q_{\perp m}^{i}$ are defined with respect to the line from
the junction to the quarks, they depend implicitly on $\br$.
The corresponding Hamiltonian for $\thetaj>120^o$ is obtained by
setting $l_3 = N_3 = 0$ in $\meff$ and $\beff$ in Eqs.~\ref{meff}
and~\ref{beff}, and restricting the summation in Eq.~\ref{ham2} to
$i=1,2$, with $\vbead$ given by Eq.~\ref{vbead1}.

Since Eq.~\ref{ham2} is diagonal in the coordinates of the beads, the
last two terms in Eq.~\ref{ham2}, corresponding to the bead
energies, can be replaced by their ground-state harmonic oscillator
energies
\beqn 
\plabel{ham}
{\tilde H}_{\mbox{\small flux}}=\frac{1}{2}\meff{\bf \dot{r}^2}+\vbead
+\sum_{i=1}^3 \sqrt{\frac{b l_i}{\beff(N_i+1)}} 
\sum_{m=1}^{N_i} \omega^i_m,
\eeqn
which is summed over the two polarizations possible for each bead
vibration. The sum of the frequencies is
\beqn
\sum_{m=1}^{N_i}\omega^i_m 
= \sqrt{2} \frac{N_i+1}{l_i} \frac{\sin \frac{N_i\pi}{4(N_i+1)}}
{\sin \frac{\pi}{4(N_i+1)}}.
\eeqn
If the number of beads is taken to infinity, the dependence of the
Hamiltonian on the (unphysical) number of beads can be removed. The
part of the Hamiltonian in Eq.~\ref{ham} arising from the beads
becomes infinite when $N_i \rightarrow \infty$. To avoid infinite
energies, the bead separation regularization parameter (analogous to
the lattice spacing) $a\equiv l_i/(N_i+1)$ is fixed when
$N_i \rightarrow \infty$. This is consistent with the flux-tube model
philosophy that $a$ cannot be chosen arbitrarily small, since that
would lead to the breakdown of the strong coupling expansion of the
Hamiltonian formulation of QCD, from which the model is
motivated~\cite{paton85}. Taking $N_i \rightarrow \infty$, while
keeping $a$ fixed, the Hamiltonian becomes~\cite{paton85}
\beqn \plabel{ham3}
{\tilde H}_{\mbox{\small flux}}^{\infty}=
\frac{1}{2}\meff^{\infty}{\bf \dot{r}^2}+\vbead
+\sum_{i=1}^3 \left[\frac{4l_i}{\pi a^2}-\frac{1}{a}-\frac{\pi}{12l_i} + 
{\cal O}(a^2) + \cdots\right],
\eeqn
with
\beqn\plabel{meff2}
\meff^{\infty}\equiv b\sum_{i=1}^3l_i
\left ( \frac{1}{3}- \frac{b\sum_{i=1}^3l_i}{4
\sum_{i=1}^3(bl_i+M_i)} \right ),
\eeqn 
where ${\tilde H}_{\mbox{\small flux}}^{\infty}$ is now independent of
the (unknown) ratio $m_J/m_b$.  Since $N_i \rightarrow \infty$ with
$a$ fixed, the part of the Hamiltonian in Eq. \ref{ham3} arising from
the beads is valid only in the limit $l_i \rightarrow \infty$.

The part of the Hamiltonian in Eq.~\ref{ham3} arising from the beads
contains a linear term $4l_i/\pi a^2$ and a constant term $-1/a$,
which are regularization-scheme dependent terms in the ``self energy''
of the string system. As explained in Ref.~\cite{paton85}, the linear
term should be regarded as a contribution that renormalizes the bare
string tension $b$ to its physical value. The constant term is three
times larger than the constant term found for mesons~\cite{paton85}.
The Luscher~\cite{luscher} term $-\pi/12l_i$ is regularization-scheme
independent and finite, and can be regarded as a prediction of the
flux-tube model, although it is insignificant at large $l_i$, where
its derivation is valid. The Luscher term arises in relativistic
string theories~\cite{luscher} in the limit $l_i \rightarrow
\infty$. In this limit the string excitations in our model coincide
with relativistic string theories. There exists strong lattice-QCD
evidence, through the study of ``torelons'', of the existence of the
Luscher term in QCD~\cite{stephenson}. The form of the Luscher term
should be contrasted with that of a Coulomb term, with the former
depending inversely on the triad lengths, and the latter inversely on
the distance between two quarks.

The Hamiltonian for the $\thetaj>120^o$ case is obtained by setting
$l_3 = 0$ in $\meff^{\infty}$ in Eq.~\ref{meff2}, and restricting the
summation in Eq.~\ref{ham3} to $i=1,2$ with $\vbead$ from
Eq. \ref{vbead1}.

If the redefined adiabatic approximation was not made, i.e. if the
calculation was not performed in the center of mass frame of the entire
system with the distances between the quarks fixed, then
$\meff^{\infty} = b\sum l_i/3$. The correction from center-of-mass
motion in Eq.~\ref{meff2} substantially reduces the effective mass of
the junction. It is shown below that the excitation energies of the
junction are proportional to $(\meff^{\infty})^{-\frac{1}{2}}$ (see
Eqs.~\ref{freq1}--\ref{freq2}), and so with typical QQQ configurations
the junction excitation energies are 1 to 2 times larger in the
redefined adiabatic approximation than in the adiabatic
approximation. This underlines the importance of working in the CM
frame.

The ground state bead configuration that solves the Hamiltonian of
Eq.~\ref{ham2} is
\beqna \plabel{wavphi}
\Phi &=& \lim_{N_i\rightarrow \infty}\prod^3_{i=1} \pi^{-\frac{N_i}{2}} 
\prod_{m=1}^{N_i} (\beff bl_i(N_i+1))^{\frac{1}{4}}
\sqrt{\frac{\omega^i_m}{2}}\nonumber\\
&&\exp\left\{-\frac{1}{2}\sum_{i=1}^3
\sum_{m=1}^{N_i} \sqrt{\beff bl_i(N_i+1)}\,\frac{\omega^i_m}{2} 
\left[(q_{\parallel m}^{i})^2+(q_{\perp m}^{i})^2\right] \right\},
\eeqna
in the limit $N_i \rightarrow \infty$ (with $a$ fixed) that is used to
express the Hamiltonian in Eq.~\ref{ham3}.  The corresponding wave
function for $\thetaj>120^o$ is obtained by restricting the products
in Eq.~\ref{wavphi} to $i=1,2$, and setting $l_3 = 0$ in $\beff$
in Eq.~\ref{beff}.

\subsection{Analytic small-oscillations solution to the junction 
Hamiltonian for $\thetaj= 120^o$ \plabel{analy}}
Define
\beqn\plabel{rho}
{\brho}\equiv \frac{\br_1 - \br_2}{\sqrt{2}}\hspace{1cm}
{\blambda} \equiv \frac{\br_1+\br_2-2 \br_3}{\sqrt{6}}\hspace{1cm}
\cos \thetarl \equiv \frac{\brho\cdot\blambda}{\rho\lambda}.
\eeqn
The six Jacobi variables $\brho$, $\blambda$ consist of (1) four
variables which specify the positions of the quarks in the $QQQ$
plane, $\rho\equiv|\brho|$, $\lambda\equiv|\blambda|$, and $\thetarl$
(the angle between $\brho$ and $\blambda$) and $\phirho$ (the angle
between $\brho$ and the space-fixed $x$-direction), and (2) two polar
angles $\theta$, and $\phi$ which specify the orientation of the
vector $\brho\times\blambda$ which lies perpendicular to the plane.
The variables $\rho, \lambda$ and $\thetarl$ are rotational
scalars. They are related to the triad lengths $l_1$, $l_2$, and $l_3$
by the relations
\begin{eqnarray}
\label{Jacobi}
l_1 = {\cal N}\; (\rho^2+\rho\lambda\sin\thetarl+
\sqrt{3}\rho\lambda\cos\thetarl) &&
l_2 = {\cal N}\; (\rho^2+\rho\lambda\sin\thetarl-
\sqrt{3}\rho\lambda\cos\thetarl) \nonumber \\
l_3 = {\cal N}\; (-\frac{\rho^2}{2}+\frac{3}{2}\lambda^2+
\rho\lambda\sin\thetarl) &&
{\cal N}^{-1} = \sqrt{\frac{3}{2} (\rho^2+\lambda^2) 
+ 3 \rho\lambda\sin\thetarl}.
\end{eqnarray}

In the remainder of this section junction motion is treated in the
small-oscillations approximation for the $\thetaj=120^o$ case.  The
junction Hamiltonian is $\frac{1}{2}\meff^{\infty}{\bf
\dot{r}^2}+\vbead^{\mbox{\small sm.osc.}}$, where the potential is
from Eq.~\ref{vbead3}. Junction motion in $(x,y)$ and out ($z$) of
the QQQ plane are not coupled in $\vbead^{\mbox{\small sm.osc.}}$, so
motion along the $z$ direction is one of the vibrational modes of the
junction. One way to define the $z$ direction in terms of the
positions of the quarks is by the normalized vector
\beqn \plabel{etahat}
\etahat_z =  \sigma_z \frac{{\brho}\times{\blambda}}
{|{\brho}\times{\blambda}|},
\eeqn
where $\sigma_z$ denotes a sign that will be specified later. Note that
$z$ motion of the junction motion occurs along the direction of the
vector ${\brho}\times{\blambda}$, but there is no physical reason to
prefer one sign $\sigma_z$ over the other.

The in-plane part of the Hamiltonian can be diagonalized in terms of
the normalized eigenvectors~\footnote{The expression in
Eq.~\ref{etahat1} for $\etahat_{\pm}$ for $l_1=l_2=l_3$ and the
expression for $\etahat_{-}$ for $l_2=l_3$ are not well-defined.}
\beqn \plabel{etahat1}
\etahat_{\pm} = \frac{\sigma_\pm}{N(l_1,l_2,l_3)} \left\{\left[
\frac{1}{l_3}-\frac{1}{2}(\frac{1}{l_1}+\frac{1}{l_2})\pm 
\sqrt{s(l_1,l_2,l_3)}\right]\hat{\bf x}
+\frac{\sqrt{3}}{2}\{ \frac{1}{l_2}-\frac{1}{l_1} \}\hat{\bf y}\right\},
\eeqn
where 
\beqn
\plabel{sli}
s(l_1,l_2,l_3)\equiv \frac{1}{l_1^2}+\frac{1}{l_2^2}+\frac{1}{l_3^2}
-\frac{1}{l_1l_2}-\frac{1}{l_1l_3}-\frac{1}{l_2l_3}>0,
\eeqn
and
\beqn N(l_1,l_2,l_3)\equiv
\sqrt{2}\sqrt{s(l_1,l_2,l_3)\pm
(\frac{1}{l_3}-\frac{1}{2}(\frac{1}{l_1}+\frac{1}{l_2}))
\sqrt{s(l_1,l_2,l_3)}},
\eeqn
and a sign $\sigma_\pm$ is included for the same reasons as above. The
vectors $\etahat_z$, $\etahat_+$, and $\etahat_-$ can be verified to
be orthonormal vectors. The junction Hamiltonian can now be written as
\beqna \plabel{ham4}
\lefteqn{\frac{1}{2}\meff^{\infty}{\bf \dot{r}^2}
+\vbead^{\mbox{\small sm.osc.}}=
b\sum_{i=1}^3 l_i + \nonumber }  \\ 
& &
\frac{1}{2}\meff^{\infty}\left[ (\etahat_+\cdot {\bf \dot{r}})^2+
(\etahat_-\cdot {\bf \dot{r}})^2+(\etahat_z\cdot {\bf \dot{r}})^2
+\omega^2_+(\etahat_+\cdot\br)^2+
\omega^2_-(\etahat_-\cdot\br)^2+\omega^2_z(\etahat_z\cdot\br)^2
\right],
\eeqna
where the vibrational frequencies are given by
\beqn \plabel{freq1}
\omega^2_z = \frac{b}{\meff^{\infty}}(\frac{1}{l_1}+
\frac{1}{l_2}+\frac{1}{l_3})
\eeqn
and
\beqn \plabel{freq2} \omega^2_{\pm}=\frac{b}{2\meff^{\infty}}\left
(\frac{1}{l_1} +\frac{1}{l_2}+\frac{1}{l_3}\pm
\sqrt{s(l_1,l_2,l_3)}\right ).  
\eeqn 
Note that the out-of-plane mode is always more energetic than the
in-plane modes, since $\omega_z > \omega_{\pm}$, and that the in-plane
modes have $\omega_- \leq \omega_+$, with degeneracy only when
$l_1=l_2=l_3$.

Solving the Schr\"{o}dinger equation corresponding to Eqs.~\ref{ham3}
and~\ref{ham4} yields the ground-state energy, corresponding to the
adiabatic potential for the quark motion in a conventional baryon, of
\beqn \plabel{ad} 
V_B (l_1,l_2,l_3) = b\sum_{i=1}^3 l_i + 
\frac{1}{2}(\omega_+ +\omega_- +\omega_z)
+\sum_{i=1}^3 \left[\frac{4l_i}{\pi a^2}-\frac{1}{a}-\frac{\pi}{12l_i}+ 
{\cal O}(a^2) + \cdots\right].
\eeqn
Junction excitations in the $\etahat_-$, $\etahat_+$, or $\etahat_z$
directions yield adiabatic potentials for the quark motion in
different low-lying hybrid baryons, denoted $H_1$, $H_2$, and $H_3$,
ordered from least to most energetic. The hybrid baryon string energy
(adiabatic potential) is that of the baryon in Eq.~\ref{ad} with the
term $\omega_-$, $\omega_+$ or $\omega_z$ added for $H_1$, $H_2$ or $
H_3$ hybrid baryons, respectively. Note that these results neglect the
junction-bead and bead-bead interactions, which has been demonstrated
to be a good approximation only for the lowest energy ($\omega_-$)
mode.

It is intriguing to note that the baryon potential in Eq.~\ref{ad}
serves as an analytical form to which the infinitely-heavy quark
potentials calculated in lattice QCD can be fitted as a function of
$l_i$.  Furthermore, the $l_i$ dependence of the various hybrid baryon
potentials are predicted.  Potentials were also predicted in
Ref.~\cite{kalash}.  Comparisons to lattice results would be
instructive at large $l_i$, for which Eq.~\ref{ad} was derived. The
physical string tension is $b-4/(\pi a^2)$. The constant $-1/a$ term
is regularization-scheme dependent, and hence not physical. Indeed,
lattice calculations~\cite{bali} find the constant term regularization
dependent, and proportional to $1/a$. The remaining terms do not
depend on either $b$ or $a$ when $M_i\rightarrow\infty$, which is the
limit in which lattice QCD potentials are evaluated, noting that
$b/\meff^\infty$ is independent of $b$ in this limit.

The normalized flux wave function of the baryon is
\beqn \plabel{psib}
\Psi_B(\br)=
\frac{(\meff^{\infty})^{\frac{3}{4}}
(\omega_+\omega_-\omega_z)^{\frac{1}{4}}}{\pi^{\frac{3}{4}}}
\exp\left\{ - \frac{\meff^{\infty}}{2}(\omega_+(\etahat_+\cdot\br)^2+
\omega_-(\etahat_-\cdot\br)^2+\omega_z(\etahat_z\cdot\br)^2)\right\} \;\Phi,
\eeqn
where the bead wave function $\Phi$ in Eq.~\ref{wavphi} has been
incorporated. For the $H_1$, $H_2$, and $H_3$ hybrid baryons,
respectively, the normalized flux wave functions are
$\Psi_{H_i}(\br)$
\beqn \plabel{psih}
\sqrt{2 \meff^{\infty} \omega_-}\;\etahat_-\cdot\br \;\Psi_B(\br),\hspace{1cm}
\sqrt{2 \meff^{\infty} \omega_+}\;\etahat_+\cdot\br \;\Psi_B(\br),\hspace{1cm}
\sqrt{2 \meff^{\infty} \omega_z}\;\etahat_z\cdot\br \;\Psi_B(\br).
\eeqn

\subsection{Junction motion away from the small-oscillations limit\plabel{num}}
Eq. \ref{vbead1} cannot be expanded in the small-oscillations
approximation to the junction motion. Without this approximation,
i.e. where $\vbead$ in ${\tilde H}_{\mbox{\small flux}}^{\infty}$ in
Eq. \ref{ham3} is taken from Eqs. \ref{vbead2} and \ref{vbead1} for
the cases $\thetaj=120^o$ and $\thetaj>120^o$ respectively, the
eigenfrequencies and eigenvectors cannot be solved for analytically.

The variational principle is used to separately minimize the
expectation value of the Hamiltonian ${\tilde H}_{\mbox{\small
flux}}^{\infty}$ by solving the Schr\"odinger equation for the
conventional baryons and hybrids $H_i$ using the ansatz
simple-harmonic-oscillator wave functions in Eqs.~\ref{psib}
and~\ref{psih}.  The calculated energies are upper bounds for the true
energies, according to the Hyleraas-Undheim theorem. The parameters of
the ansatz wave functions no longer have the values that they had in
the small-oscillations approximation, but need to be fitted. For
example, the directions $\etahat_z$, $\etahat_-$, and $\etahat_+$ are
no longer given by Eqs.~\ref{etahat} and~\ref{etahat1}, but will be
fixed by the variational principle.

Note that ${\tilde H}_{\mbox{\small flux}}^{\infty}$ (Eq.~\ref{ham3})
is even under the discrete transformation $z\rightarrow -z$ since
$\vbead$ in Eqs.~\ref{vbead2} and~\ref{vbead1} only depends on
$z^2$. This implies that the wave functions should be either odd or
even under $z\rightarrow -z$. But since the wave functions are assumed
to be of the form in Eqs.~\ref{psib} and~\ref{psih}, it is not
difficult to show that this implies that one of the junction
vibrational modes, corresponding to $\etahat_z$, is always
perpendicular to the $QQQ$ plane.  Note that $\etahat_{-}$,
$\etahat_{+}$, and $\etahat_z$ are required to be orthonormal in order
to obtain orthonormal hybrid baryon wave functions in Eq.~\ref{psih}.
This gives four variational parameters that specify the ansatz
wave functions: $\meff^{\infty}\omega_-$, $\meff^{\infty}\omega_+$,
$\meff^{\infty}\omega_z$, and an angle that describes the ray in which
$\etahat_-$ lies in the plane relative to the $x$-direction defined
in Fig.~\ref{hyb_coords}. The minimization is carried out with respect
to these four variables.

\section{Flux Symmetry\label{fluxsymmetry}}

\subsection{Quark label exchange symmetry}
Denote by $P_{12}$, $P_{13}$, and $P_{23}$ the permutations which
exchange the labels of the quarks. Except for color, quark-spin and flavor
labels, which will only be of interest later, exchange symmetry
affects only position labels.  Under such quark label permutations the
positions of the quarks are exchanged, e.g. $P_{12}$ exchanges $\br_1
\leftrightarrow \br_2$, but note that variables that are not functions
of the $\br_i$ are unaffected. Since the physics does not depend on
the quark position labelling convention, the flux Hamiltonian given by
Eq.~\ref{ham1} should be exchange symmetric. As the equilibrium
junction position $\req$ in Eq.~\ref{req} is invariant under the
$P_{ij}$, and $\bl_i = \br_i - \req$, it follows that $\bl_i
\leftrightarrow \bl_j$ under $P_{ij}$.  Also, since the number of
beads on the $i$-th triad is $N_i = l_i/a-1$, then $N_i
\leftrightarrow N_j$ under $P_{ij}$.  The potential $\vbead$ in
Eq.~\ref{vbead2} can be written in the manifestly exchange-symmetric
form
\beqn 
\vbead = b\sum_{i=1}^3 |\br_i-\br|,
\eeqn
noting that the junction position $\br$ is not determined by the
positions of the quarks. This establishes that all quantities in the
flux Hamiltonian in Eq.~\ref{ham1} for the $\thetaj=120^o$ case are
invariant under exchange symmetry transformations.

Since the flux Hamiltonian is invariant under exchange symmetry,
it is clear that energies (or adiabatic potentials) that are solutions
of the flux Schr\"{o}dinger equation are also 
exchange symmetric. This is explicit for the
frequencies in Eqs.~\ref{freq1} and~\ref{freq2}, and the potential
in Eq.~\ref{ad}.

By the same arguments as above it can be shown that in the
$\thetaj>120^o$ case, the Hamiltonian (Eq.~\ref{V120}) is invariant
under $P_{12}$.

Since the Hamiltonian is exchange symmetric, the commutation
relations $[H_{\mbox{\small flux}},P_{ij}]=0$ hold. This implies that the
wave functions of conventional and hybrid baryons have to represent the
permutation group $S_3$. Possible representations are the
one-dimensional symmetric and antisymmetric representations, and the
two-dimensional mixed-symmetry representation. Since the baryon and
each of the hybrid baryons $H_i$ have different~\footnote{Except when
$l_1=l_2=l_3$.} flux energies $V(l_1,l_2,l_3)$, where $H_{\mbox{\small
flux}}\Psi = V(l_1,l_2,l_3)\Psi$, each of the four wavefunctions
$\Psi_B(\br)$ and $\Psi_{H_i}(\br)$ have to belong to a one
dimensional representation, as they cannot mix with each other under
any of the permutations $P_{ij}$. This implies that $\Psi_B(\br)$ and
$\Psi_{H_i}(\br)$ are either totally symmetric or antisymmetric under
quark label exchange.

In the baryon wave function of Eq.~\ref{psib}, the quantities
$\omega_+$, $\omega_-$, $\omega_z$, and $\meff^{\infty}$ are exchange
symmetric, so that the factor before the exponential is invariant. The
bead wave function $\Phi$, given in Eq.~\ref{wavphi}, is also
invariant. Since $\Psi_B(\br)$ is either exchange symmetric or
antisymmetric, the exponential function in the junction coordinates
must be either exchange symmetric or antisymmetric. The second
possibility is untenable since the exponential function is always
positive.  Hence, the baryon wave function {\it $\Psi_B(\br)$ is
totally symmetric under exchange symmetry.}

Consider the hybrid baryon wavefunctions in Eq.~\ref{psih}. The above
implies that $\etahat_-$, $\etahat_+$, and $\etahat_z$ are either
totally symmetric or totally antisymmetric under exchange symmetry,
since $\br$ is independent of the quark labels. It is shown in
Appendix~\ref{app2} that both possibilities are explicitly realizable.
This implies that for each of the hybrid baryons $H_i$, {\it there is a
degenerate pair of totally symmetric (S) and totally antisymmetric
(A) wave functions}, denoted by $H_i^S$ and $H_i^A$.

The preceding argument assumed that $\Psi$ has the form in
Eqs.~\ref{psib} and~\ref{psih}, which applies only in the
$\thetaj=120^o$ case with small junction oscillations. However,
as was discussed in section~\ref{num}, in the more general case where
small junction oscillations are not assumed, ansatz wave
functions of the form in Eqs.~\ref{psib} and~\ref{psih} are used, so that
the preceding arguments regarding exchange symmetry remain valid.

The above arguments can be repeated to show that in the
$\thetaj>120^o$ case, the baryon ansatz wave function $\Psi_B(\br)$ is
invariant under $P_{12}$, and that the hybrid baryon ansatz wave
functions $\Psi_{H_i}(\br)$ and $\etahat_{-}$, $\etahat_{+}$, and
$\etahat_z$ are either odd or even under $P_{12}$.

\subsection{Parity \label{par}} 

The operation of the inversion of all coordinates, or parity, is a
symmetry of the flux Hamiltonian. It follows that $\etahat_z$ in
Eq.~\ref{etahat} is {\it even under parity}, since $\brho\rightarrow
-\brho$ and $\blambda \rightarrow -\blambda$ under parity. If
$\sigma_z=1$ this follows trivially, and if $\sigma_z$ is given by
Eq.~\ref{eps}, it follows because Eq.~\ref{eps} is invariant under
parity.

The $l_i$ remain invariant under parity since they are lengths, but
the $\hat{\lpmb{l}}_i$ are odd under parity.  From the definition of
the $\etahat_{\pm}$ in Eq.~\ref{etahat1}, and the definition of
$\hat{\bf x}$ and $\hat{\bf y}$ in terms of the $\hat{\lpmb{l}}_i$ in
Eq. \ref{xhat}, it follows that the $\etahat_{\pm}$ are {\it odd under
parity}. The sign $\sigma_\pm$ is invariant under parity (see
Appendix~\ref{app2}). This argument is so far valid only when
$\etahat_{\pm}$ is given by Eq. \ref{etahat1}, applicable for the
$\theta_J=120^{\rm o}$ case in the small-oscillations
approximation. However, for the ansatz variational wave functions in
section \ref{num}, the $\etahat_{\pm}$ lie in the $QQQ$ plane and so
must be linear combinations of $\brho$ and $\blambda$ with
coefficients which are functions of the parity-invariant variables
$\rho$, $\lambda$ and $\thetarl$, so the $\etahat_{\pm}$ remain odd
under parity.

Since the position $\br$ of the junction is a vector, it is odd
under parity. It follows that the baryon wave function $\Psi_B(\br)$
in Eq.~\ref{psib} is invariant under parity.  The hybrid baryon wave
functions in the $QQQ$ plane, i.e. $\Psi_{H_1}(\br)$ and
$\Psi_{H_2}(\br)$ in Eq.~\ref{psih}, are even under parity, while
$\Psi_{H_3}(\br)$ is odd under parity. These results also obtain for
$\theta_J>120^{\rm o}$.

In summary, flux wave functions of baryons and $H_{1,2}$ hybrid
baryons are even under parity, while the $H_3$ hybrid baryon flux wave
functions are odd under parity.

\subsection{Chirality \plabel{chi}}

Reflection in the QQQ plane, or ``chirality''~\cite{morningstar97}, is
generally a symmetry of the flux wave function in the adiabatic
approximation, since the physics does not distinguish between above
and below the $QQQ$ plane. The relevant group consists of the identity
and reflection transformations. In this approximation the flux wave
function can be classified according to its eigenvalue under
reflections in the plane spanned by the three quarks, which is the
chirality $\pm 1$.

In the flux-tube model this reflection takes $z\rightarrow
-z$ and $q_{\perp m}^{i}\rightarrow -q_{\perp m}^{i}$. The most
general Hamiltonian derived in this work, Eq.~\ref{ham1}, is invariant
under this reflection transformation, as it must be. The baryon and
hybrid baryon wave functions in Eqs.~\ref{psib} and~\ref{psih} are 
eigenfunctions of the reflection transformation. The baryon and
``planar'' hybrids ($H_{1,2}$) have chirality 1, and the
``non-planar'' hybrid $H_3$ has chirality -1. Hence the chirality
formally allows us to clearly distinguish ``planar'' and
``non-planar'' hybrids, even for more general solutions of the
Hamiltonian than those in Eqns.~\ref{psib} and~\ref{psih}.

In adiabatic lattice QCD, exchange symmetry, parity and chirality
should classify the (hybrid) baryon flux wave functions. These
properties are sometimes called the ``quantum numbers of the adiabatic
surface''.

\section{Quantum numbers\label{qmnos}}

\subsection{Orbital Angular Momentum \plabel{hil}}

For every set of quark positions ${\bf r}_i$ the potential in which
the junction moves is anisotropic, which means that the solutions of
the Schr\"odinger equation for the junction motion do not have
definite orbital angular momentum or its projection. However, in the
absence of the adiabatic approximation the combined wavefunction of
the quark and junction motions must be a state of good angular
momentum.

It is possible to determine the angular momentum character of the
variational wavefunctions which minimize the flux energy for a given
set of quark positions. The probability of overlap between an
isotropic $S$-wave harmonic-oscillator state with frequency $\omega$
and the baryon flux wavefunction $\Psi_B(\br)$ of Eq.~\ref{psib} is
\beqn
\label{P0}
P_0(l=0)\equiv \left\vert \left\langle \Psi_B(\br) \vert 000\right\rangle\right\vert^2
=\frac{8 \sqrt{\omega^3 \omega_- \omega_+ \omega_z}}
{(\omega_- + \omega)(\omega_+ + \omega)(\omega_z + \omega)},
\eeqn
and that of an isotropic $P$-wave harmonic-oscillator state with the
flux wavefunction of the lightest hybrid $\Psi_{H_1}(\br)$ of
Eq.~\ref{psih} is
\beqn
\label{P1}
P_1(l=1)\equiv \sum_{M=-1,+1} \left\vert \left\langle \Psi_{H_1}(\br) \vert 01M\right\rangle
\right\vert^2
=\frac{32 \sqrt{\omega^5 \omega_-^3 \omega_+ \omega_z}}
{(\omega_- + \omega)^3(\omega_+ + \omega)(\omega_z + \omega)}.
\eeqn
Once the energies of the ground and first excited states of the flux
have been independently minimized in the variational calculation
described below in Sec.~\ref{fluxpot}, the calculated values of
$\omega_-$, $\omega_+$, and $\omega_z$ can be used to find these
probabilities. The result of these numerical studies is shown for
sample quark configurations in Table~\ref{fluxenergy}. It is clear
that the ground state of the flux is in an almost exclusively angular
momentum zero state, so the orbital angular momentum of the baryon is
that of the quark motion.
\begin{table}[t]
\begin{center}
\caption{\small Flux energies and angular momentum probabilities
calculated using Eqs.~\protect{\ref{P0}} and~\protect{\ref{P1}} with
$\sqrt{\omega M^\infty_{\rm eff}}=0.4$ GeV, for four quark
configurations. Here $\sum_i M_i=0.99$ GeV, $b=0.18$ GeV$^2$, the
triad lengths $l_i$ are given in GeV$^{-1}$, and energies are in
GeV. Here $E_0$ and $P_0(l=0)$ are the energy and $S$-wave probability
of the flux ground state, $E_1$ and $P_1(l=1)$ are the energy and
$P$-wave probability of the first excited state of the flux.}
\vskip 0.5 cm
\plabel{fluxenergy}
\begin{tabular}{|c|c|c||c|c||c|c|}
\hline 
$l_1$ & $l_2$ & $l_3$ 
& $E_0$ & $P_0(l=0)$ 
& $E_1$ & $P_1(l=1)$\\
\hline \hline 
2.5 & 2.5 & 2.5 
& 1.09  & 0.995
& 1.76 & 0.997 \\
2.5 & 2.5 & 0.5
& 1.42 & 0.999
& 2.18 & 0.998 \\
2.5 & 5.0 & 0.5
& 1.18 & 0.993 
& 1.80 & 0.998 \\
0.5 & 0.5 & 10.0
& 1.30 & 0.986 
& 1.92 & 0.998\\
\hline 
\end{tabular}
\end{center}
\end{table}

Table~\ref{fluxenergy} shows that variational calculations result in
flux wave functions $\Psi_{H_1}({\bf r})$ which are to better than
99\% a linear combination of $Y_{11}(\hat{\bf r})$ and
$Y_{1-1}(\hat{\bf r})$. An alternative argument is given here that the
angular momentum of the flux in the lowest-lying hybrid baryons
($H_1$) is predominantly unity.  The flux wave function in
Eq.~\ref{psih} of the lightest ($H_1$) hybrid baryon is proportional
to $\hat{\lpmb{\eta}}_-\cdot{\bf r}$, where $\hat{\lpmb{\eta}}_-$ lies
in the plane of the quarks. If the exponential in Eq.~\ref{psih} was
spherically symmetric, it would be strictly true that $\Psi_{H_1}({\bf
r})$ was proportional to a linear combination of $Y_{l\,1}(\hat{\bf
r})$ and $Y_{l\,-1}(\hat{\bf r})$, with the junction position ${\bf
r}$ defined relative to a (body) $z$-axis perpendicular to the quark
plane.

Further numerical studies, described below, have shown that the
least energetic motion of the quarks in the $H_1$ adiabatic potential
has the quark angular momentum $L_q=0$, the next highest $L_q=1$,
etc. Furthermore, there is a substantial cost in energy to increase the
quark angular momentum in the $H_1$ hybrid potential, so the total
orbital angular momentum of the lightest hybrid baryon is unity.

In order for the combined flux and quark orbital angular momentum to
have a definite value (unity), in principle the components with
orbital angular momentum other than unity in the flux wavefunction
must be combined with quark motion with $L_q\geq 1$ to make the total
orbital angular momentum unity. Given the negligible size of these
components, a very good approximation to the energy can be found by
assuming that the flux orbital angular momentum is unity.

\subsection{Color \plabel{col}}

It is important to note that the wave function of the (hybrid) baryon
has both a color {\it and} a flux sector, which are separable.  This
is because color is a separable degree of freedom in quantum
chromodynamics, which labels the quarks and flux lines. This is
described by the color sector of the theory.  The flux sector, on the
other hand, concerns the dynamics of the flux. In the bag model and
large $N_c$ limit the same separation occurs, where the octet color of
the gluon is combined with that of the quarks, and the spatial motion
of the gluon is treated separately~\cite{bag,yan}.

In the flux-tube model the color structure of a hybrid baryon is
motivated by the strong coupling limit of the Hamiltonian formulation
of lattice QCD~\cite{kogut}. Here, the quarks are sources of triplet
color, which flows along the triad connecting the quarks to the
junction, where an $\epsilon$-tensor neutralizes the color. The color
wave function is hence totally antisymmetric under exchange of quarks
for {\it both} the conventional and hybrid baryon. In the bag
model~\cite{bag} and in the large $N_c$ limit~\cite{yan} the color
structure of a hybrid baryon is very different. This color structure
is critical for the correct exchange symmetry properties of the
conventional and hybrid baryons, and hence the structure of the wave
function.

\subsection{(Hybrid) Baryon Wave Functions\label{qmno}}

The energy of the quarks in the potential given by the flux energies
is found by expanding the quark wave function in a basis with
well defined orbital angular momentum $L_q$ and projection $M_q$, made
up from orbital angular momenta $l_{\rho}$ and $l_{\lambda}$ in the
coordinates $\brho$ and $\blambda$ respectively,
\beqna
\plabel{kds} 
\langle \brho,\blambda|n_{\rho} l_{\rho}
n_{\lambda} l_{\lambda};L_q m_q\rangle & \equiv & 
{\cal N}_{n_\rho l_\rho} R_{n_{\rho} l_\rho} (\rho)\; 
{\cal N}_{n_\lambda l_\lambda} R_{n_\lambda l_\lambda}(\lambda)
\nonumber \\
&&\eqntimes \sum_{m_{\rho} m_{\lambda}}
C(l_{\rho}m_{\rho}l_{\lambda}m_{\lambda};L_q M_q)
\; Y_{l_{\rho}m_{\rho}}(\Omega_{\rho})\;
Y_{l_{\lambda}m_{\lambda}}(\Omega_{\lambda}),
\eeqna 
where ${\cal N}_{nl}$ is a normalization factor, and the
Clebsch-Gordon coefficient combines spherical harmonics with orbital
angular momentum $l_{\rho}$ and $l_{\lambda}$ to form a state with
orbital angular momentum $L_q$.  Here the $R_{nl}$ are orthonormal and
complete functions in the radial coordinate, where $n=0,1,2...$
denotes the radial quantum number, which are taken to be
three-dimensional harmonic oscillator radial wave functions,
i.e. Laguerre polynomials.  It is easy to show that the wave functions
in Eq.~\ref{kds} form an orthonormal (in all six labels) and complete
set. In Eq.~\ref{kds} a formal notation is used where the wave
function is defined as the overlap of a state $|n_{\rho} l_{\rho}
n_{\lambda} l_{\lambda};L_q M_q\rangle$, characterized by the quantum
numbers indicated, with a position state $|\brho,\blambda \rangle$.

A linear combination of the states in Eq.~\ref{kds} can be used to
form a general eigenstate of quark orbital angular momentum $L_q$ and
projection $M_q$, denoted by $|n L_q M_q \rangle$, where $n$ denotes
the radial quantum number.  The corresponding wave function is
\beqn 
\plabel{qua}
\langle \brho,\blambda|n L_q M_q \rangle \equiv
\sum_{n_{\rho} l_{\rho} n_{\lambda} l_{\lambda}} 
c^{nL_q}_{n_{\rho} l_{\rho} n_{\lambda} l_{\lambda}}\; 
\langle \brho,\blambda|n_{\rho} l_{\rho} n_{\lambda} l_{\lambda};
L_q M_q (\brho,\blambda)\rangle.
\eeqn
The coefficients in this linear combination, and the corresponding
hybrid baryon energies, are found by diagonalizing the three-quark
Hamiltonian in the basis of Eq.~\ref{kds} with the potential energy
given by the flux energy. Note that the orbital angular momentum and
spin of the quarks are good quantum numbers as the inter-quark
potential is a spatial and quark-spin scalar, even in the presence of
the Coulomb and hyperfine (spin-spin) interactions.

It has been checked numerically for the adiabatic potentials found
here that the lowest energy solutions of the Schr\"{o}dinger equation
for both conventional and hybrid baryons have $L_q=0$ quark wave
functions. In order to determine the color, flavor, quark-spin,
parity, exchange symmetry and chirality quantum numbers of these
states, it is sufficient to consider the 
$\langle \brho,\blambda |0 0 0 0 ; 0 0\rangle$ 
component of the $L_q=0$ wave function in
Eq.~\ref{qua}, as these quantum numbers must be the same for all
components of the wave function.

From Eq.~\ref{kds}, $\langle \brho,\blambda |0 0 0 0 ; 0 0\rangle$
equals
\beqn \plabel{del}
{\cal N}_{00}^2 R_{00}(\rho)\; R_{00}(\lambda)\; Y_{00}(\Omega_{\brho})\;  
Y_{00}(\Omega_{\blambda}) = \frac{\alpha^3}{\pi^\frac{3}{2}} 
\exp \left\{-\frac{\alpha^2}{2} (\rho^2 + \lambda^2)\right\},
\eeqn
where $\alpha$ is a parameter that characterizes the Laguerre
polynomials.  This is obviously even under parity and is totally
symmetric under exchange since $\rho^2 + \lambda^2$ is exchange
symmetric. Since the parity is unaffected by the color, flavor, and
quark-spin wave functions which will multiply this $L_q=0$ spatial wave
function, the parity is determined by that of the flux wave function
given in section~\ref{par}. The parities of the low-lying hybrids are
displayed in Table~\ref{tabqu}.

\begin{table}
\begin{center}
\caption{\small Quantum numbers of ground state $L_q=0$ flux-tube
model (hybrid) baryons for the lowest flux-tube surfaces $B$ (the
conventional baryon) and $H_1$ (the lightest planar hybrid baryon). In
the absense of spin-dependent forces all ground states corresponding
to a given flux-tube surface [both symmetric ($S$) and antisymmetric
($A$)] are degenerate. Here $L$ is the total orbital angular momentum
of the quarks and the flux, $N$ or $\Delta$ denotes the flavor, $S$ is
the spin of the three quarks, ${\bf J} = {\bf L+S}$ is the total
angular momentum, and $P$ is the parity. Low-lying hybrid baryons in
the bag model contructed with a transverse electric gluon
(corresponding to the surfaces $H_1^S$ and $H_1^A$) are also
shown~\protect\cite{bag}.}  \plabel{tabqu}
\begin{tabular}{|c||r|l|l|}
\hline 
(Hybrid) Baryon          & Chirality & $L$ & $(N,\Delta)^{2S+1}J^P$ \\
\hline 
$B       $ & 1 & 0 &  $N^2 {\frac{1}{2}}^+, \; \Delta^4 {\frac{3}{2}}^+$\\
$H_1^S   $ & 1 & 1 &  $N^2 {\frac{1}{2}}^+, \; N^2 {\frac{3}{2}}^+, \; \Delta^4 {\frac{1}{2}}^+, \; \Delta^4 {\frac{3}{2}}^+, \; \Delta^4 {\frac{5}{2}}^+$\\
$H_1^A   $ & 1 & 1 &  $N^2 {\frac{1}{2}}^+, \; N^2 {\frac{3}{2}}^+$\\
\hline 
{\rm Bag model hybrids} &  & 1 & $N^2 {\frac{1}{2}}^+, \; N^2 {\frac{3}{2}}^+, \; N^4
{\frac{1}{2}}^+, \; N^4 {\frac{3}{2}}^+, \; N^4 {\frac{5}{2}}^+, \;
\Delta^2 {\frac{1}{2}}^+, \; \Delta^2 {\frac{3}{2}}^+$ \\
\hline 
\end{tabular}
\end{center}
\end{table}

The quark-spin ($\chi$) $\times$ flavor ($\phi$) wave function can be made
totally symmetric for quark-spin $\frac{3}{2}$ $\times$ flavor $\Delta$,
using the product of symmetric factors $\chi^S_{3/2}\phi^S_\Delta$,
and for quark-spin $\frac{1}{2}$ $\times$ flavor $N$, using the linear
combination~\cite{capt} of mixed-symmetry factors
$(\chi^{M_\rho}_{1/2}\phi^{M_\rho}_N
+\chi^{M_\lambda}_{1/2}\phi^{M_\lambda}_N)/\sqrt{2}$. It can also be
made totally antisymmetric for quark-spin $\frac{1}{2}$ $\times$ flavor $N$
using the linear combination of mixed-symmetry factors
$(\chi^{M_\rho}_{1/2}\phi^{M_\lambda}_N
-\chi^{M_\lambda}_{1/2}\phi^{M_\rho}_N)/\sqrt{2}$.

Since quarks are fermions, the combined color, space, quark-spin, flavor and
flux wave function should be totally antisymmetric under
exchange symmetry. Since for $L_q=0$ baryons and hybrid baryons the
color and space parts are totally antisymmetric and symmetric,
respectively, the flavor $\times$ quark-spin $\times$ flux part must be
totally symmetric.

For baryons the flux wave function is totally symmetric with orbital
angular momentum zero, and so their quantum numbers are exactly as
they were in the conventional quark model with an assumed static
confining potential between the quarks. As an example, the quantum
numbers of the non-strange~$L_q=0$ ground states are shown in
Table~\ref{tabqu}. The $L_q=0$ hybrid baryons $H_i^S$ have a totally
symmetric flux wave function, and so the quark-spin $\times$ flavor
structure is the same as for the corresponding $L_q=0$ baryons,
i.e. the symmetric products above with quark-spin 1/2 for nucleons,
and quark-spin 3/2 for $\Delta$ states. For $L_q=0$ hybrid baryons
with a totally antisymmetric flux wave function, $H_i^A$, the quark-spin
$\times$ flavor wave function must be totally antisymmetric, and the
only possibility is the antisymmetric product with nucleon flavor and
quark-spin 1/2, as shown in Table~\ref{tabqu}.

Chirality is a reflection in the $QQQ$ plane, and hence only affects
the flux part of the wave function, so that its values are those given
in section~\ref{chi}.

For the lightest $L_q=0$ (hybrid) baryons the total orbital angular
momentum is that of the flux. This gives $L=0$ for low-lying
conventional baryons, so that $J=S$. Since $L=1$ for the low-lying
$H_1$ hybrid baryons, $J=\frac{1}{2}$ or $\frac{3}{2}$ for
$S=\frac{1}{2}$, and $J=\frac{1}{2}$ ,$\frac{3}{2}$, or $\frac{5}{2}$
for $S=\frac{3}{2}$, as shown in Table~\ref{tabqu}.

\section{Hamiltonian for the quark motion\label{fluxpot}}

A phenomenological form is used here for the quark Hamiltonian which
is fit to conventional baryon spectroscopy in
Ref.~\cite{capstick86}. In the case of hybrid baryons the difference
between the adiabatic potential found from numerical calculation of
the energy of the ground state and the first excited state
of the flux is added to the quark Hamiltonian.

The quark Hamiltonian has the form
\beqn \plabel{pot1}
H^{qqq} = \sum_{i=1}^3 \sqrt{{\bf P}_i^2+M_i^2} 
+\sum_{i<j}V^{\rm Coul}_{ij} +\sum_{i<j}V^{\rm cont}_{ij} 
+ {\bar V}(l_1,l_2,l_3),
\eeqn
where ${\bf P}_i$ is the momentum operator of the $i$-th quark, $M_i =
0.22$ GeV for light quarks, $b=0.18$ GeV$^2$, and the Coulomb
potential $V^{\rm Coul}_{ij}$ and hyperfine contact potential $V^{\rm
cont}_{ij}$ have the same form as in Ref.~\cite{capstick86}. The
justification for adopting this form of the Coulomb and hyperfine
contact interaction is outlined in Sec.~\ref{coul} below. For the
conventional baryon the adiabatic potential ${\bar V}_B(l_1,l_2,l_3)$
also has the form $b\sum_i l_i$ adopted in Ref.~\cite{capstick86}. In
what follows the numerical calculation of the form of the adiabatic
potential ${\bar V}$ for the lightest hybrid baryon is outlined.

\subsection{Numerical adiabatic potentials\plabel{numadpot}}

As it is not possible in all quark configurations to derive the
adiabatic potential of baryons and hybrid baryons in the
small-oscillations approximation to the junction motion given by
Eq.~\ref{ad}, a numerical calculation is used to find the flux energy,
which is part of the potential energy for the quark motion, for all
quark configurations. As discussed previously, the linear term in
Eq.~\ref{ad} which arises from the bead motion is
regularization-scheme dependent, and can be absorbed into the physical
linear term in the potential. Also, there will be no need to consider
constant and Luscher terms in this section as they are identical for
the conventional and hybrid baryons.

\begin{figure}
\includegraphics[width=10cm,angle=0]{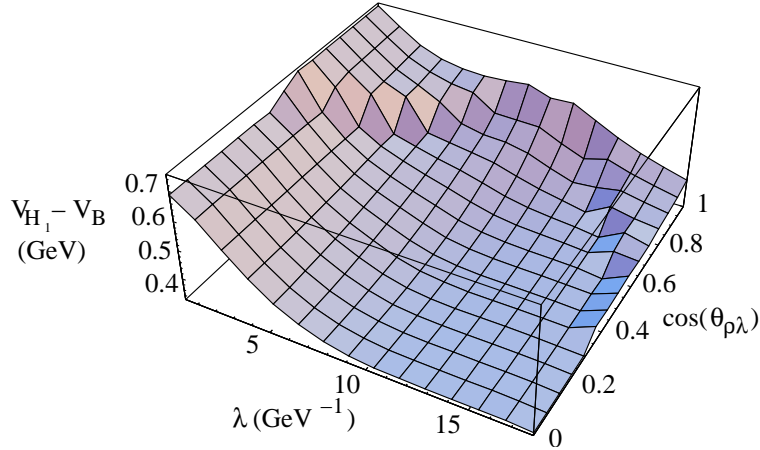}
\caption{\label{potential1} Difference $V_{H_1}-V_B$ of the hybrid and
conventional baryon adiabatic potentials for $\rho = 6.2$ GeV$^{-1}$,
as a function of $\lambda$ and $\thetarl$.}
\end{figure}
The procedure of numerically evaluating the hybrid baryon potential is
as follows.  For a large set of quark configurations
$\{l_1,l_2,l_3\}$, the Schr\"{o}dinger equation
\beqn
\label{Scheqn}
\left(\frac{1}{2}\meff^{\infty}{\bf
\dot{r}^2}+\vbead\right)\Psi(\br) =V(l_1,l_2,l_3)\Psi(\br)
\eeqn
is solved variationally for the wave function $\Psi_B(\br)$ and energy
$V_B(l_1,l_2,l_3)$ of the ground state of the junction Hamiltonian, as
described in section \ref{num}, using the ansatz wave function in
Eq.~\ref{psib}. The lowest-lying hybrid ($H_1$) baryon potential is
found by solving Eq.~\ref{Scheqn} variationally for the wave function
$\Psi_{H_1}(\br)$ and energy $V_{H_1}(l_1,l_2,l_3)$ of the first excited
state of the junction Hamiltonian, using the first ansatz wave
function in Eq.~\ref{psih}. This corresponds to junction motion in the
$QQQ$ plane, and is used because the analytical solutions in
section~\ref{analy} suggest that the lowest hybrid baryon energy can
be described by such junction motion.  The minimizations for the
baryon and lightest hybrid baryon potentials are carried out
independently. It has been checked that the numerical potentials
calculated here in the long-string limit agree with the analytic
expressions derived in that limit in Sec.~\ref{analy}, to within 2\%.
\begin{figure}
\includegraphics[width=10cm,angle=0]{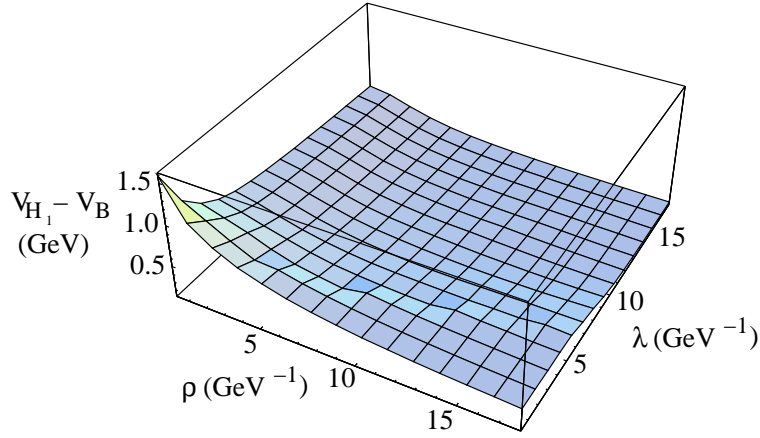}
\caption{\label{potential2} Difference $V_{H_1}-V_B$ of the hybrid and
conventional baryon adiabatic potentials for $\thetarl = \pi/2$ as a
function of $\rho$ and $\lambda$.}
\end{figure}

\begin{figure}
\includegraphics[width=10cm,angle=0]{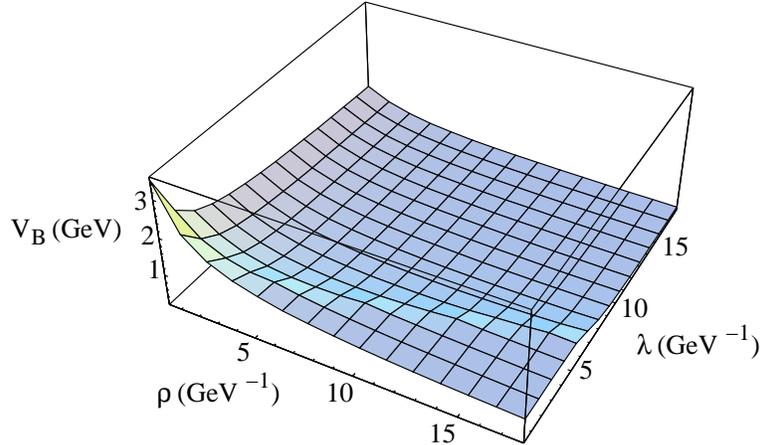}
\caption{\label{potential3} Conventional baryon adiabatic potential
without the confining potential, $V_B-b\sum_i l_i$, for $\thetarl =
\pi/2$, as a function of $\rho$ and $\lambda$.}
\end{figure}

The hybrid baryon adiabatic potential is defined to be
\beqna\plabel{pot2}
\mbox{\={V}}_{H_1} (l_1,l_2,l_3) &\equiv&
\mbox{\={V}}_B (l_1,l_2,l_3) + V_{H_1} (l_1,l_2,l_3) - V_B (l_1,l_2,l_3)
\nonumber\\
&=&b\sum_i l_i  + V_{H_1} (l_1,l_2,l_3) - V_B (l_1,l_2,l_3).
\eeqna
Selected numerical results for the difference $V_{H_1}-V_B$ of the hybrid
and conventional baryon adiabatic potentials are plotted in
Figures~\ref{potential1}--\ref{potential2}. In Fig.~\ref{potential1}
the potential is plotted with fixed $\rho$ (proportional to the
separation of quarks 1 and 2) and variable $\lambda$ (proportional to
the separation of the center-of-mass of quarks 1 and 2, and quark 3)
and $\thetarl$ (the angle between the vectors $\lpmb{\rho}$ and
$\lpmb{\lambda}$), which clearly demonstrates a discontinuity in the
derivative when the flux goes from the shape with $\theta_J=120^o$ to
that with $\theta_J>120^o$. In Figs.~\ref{potential2}
and~\ref{potential3} the behavior of $V_{H_1}-V_B$ and $V_B-b\sum_i l_i$
when $\lpmb{\rho}$ and $\lpmb{\lambda}$ are orthogonal is plotted
against $\rho$ and $\lambda$. It is obvious that both the conventional
and hybrid baryon adiabatic potentials increase when $\rho\lambda$ is
small, with the hybrid adiabatic potential increasing faster.  If the
small-oscillations approximation were employed they would tend to
infinity as $\rho\lambda\rightarrow 0$. Solving for the energy
variationally has softened this behaviour considerably.

The value of $\sqrt{\meff^{\infty}\omega_{-}}$ for the baryon is
approximately in the range $0.37 - 0.5$ GeV, while the hybrid baryon
is $\approx 0.35 - 0.48$ GeV.

\subsection{Short-distance interactions between the quarks\plabel{coul}}

There are two important configurations of the quarks when considering
the Coulomb interactions. These are where two quarks are near to each
other and the third is distant (meson-like configurations), and where
all three quarks are close to each other. It is possible to focus on
the former, because the latter is atypical and contributes little to
the energy of the baryon.

In the flux-tube model, in the meson-like configuration a string
extends from the distant quark to the other two quarks, i.e. the
system looks like a meson with the two nearby quarks in color $\bf
\bar{3}$. The reason for this is that the long-distance picture is
still appropriate for the distant quark, which means that its color
flows along the long string, and must be cancelled by the two nearby
quarks. The two nearby quarks are hence in color $\bf \bar{3}$ for
both the conventional and hybrid baryon.  This implies that the
Coulomb interactions between the nearby quarks are attractive, and
identical in the conventional and hybrid baryon.

This also has implications for hyperfine contact (spin-spin)
interactions, the existence of which has recently been confirmed in
lattice QCD~\cite{karsh}, since their form is given by the color
representation of the interacting quark pair. The interactions are,
therefore, identical for conventional and hybrid baryons. As seen in
Table~\ref{tabqu} above, the lightest $\Delta$ hybrid baryons have the
same quark-spin structure as the ground-state $\Delta$, and both the
symmetric and antisymmetric lightest nucleon hybrids have the same
quark-spin structure as the ground-state nucleon. This implies that
the lightest $\Delta$ hybrid baryons are always heavier than the
lightest $N$ hybrids, due to the same hyperfine interaction which
makes the $\Delta$ heavier than the nucleon.

\section{Numerical mass estimate for low-lying hybrid baryons\label{numerical}}

The Hamiltonians in Eqs.~\ref{pot1} and~\ref{pot2} are evaluated using
the basis of coupled three-dimensional harmonic oscillator wave
functions in Eq.~\ref{kds}, expanded up to at least the $N=7$
oscillator level, where
$N=2(n_\rho+n_\lambda)+l_\rho+l_\lambda$. These matrices are
subsequently diagonalized to yield the energies. The resulting full
wave functions (Eq.~\ref{qua}), which are linear combinations of the
harmonic oscillator wave functions, are solutions of the
Schr\"{o}dinger equation for the quark motion in the presence of the
usual ${\bar V}_B$ baryon confining potential and the ${\bar V}_{H_1}$
hybrid-baryon adiabatic potential.  The differences between the
energies for the hybrid and the conventional baryon are then added to
the experimental mass of the lightest baryon (the nucleon).

It is interesting to determine what sets the scale of the energy
difference between the lightest hybrid and conventional baryons in
this model. This can be illustrated by examining the analytic solution
to the junction Hamiltonian developed above, in the limit of small
oscillations and a large number of beads. Equation~\ref{freq2} gives
the frequency $\omega_-$ of the lowest energy string excitation in
terms of the string tension $b$, the effective mass $M^\infty_{\rm
eff}$ of the junction in the limit of a large number of beads, and the
lengths $l_i$ of the three lines from the rest position of the
junction to the quarks (triads). The effective mass of the junction
is, in turn, given in terms of the same quantities and the quark
masses $M_i$ in Eq.~\ref{meff2}. The scale of the energy difference is
set by the same quantities in our variational calculations, which do
not employ the small oscillations limit.

Consistent values of the string tension and light ($M_u=M_d$) quark
mass are used in evaluating the flux energies, which define the
adiabatic potentials in which the quarks move, and in the calculation
of the energy of the quarks (and so hybrid and conventional baryon
masses) in these potentials. The value of the string tension used is
$b= 0.18$ GeV$^2$, the typical value resulting from lattice
gauge theory calculations, which is also used in quark model
calculations of hadron masses~\cite{paton85}. The light quark masses
are set to 0.22 GeV, the same value used in Ref.~\cite{capstick86},
which also uses the relativistic kinetic energy, and Coulomb and
hyperfine contact potentials identical to those in the current
calculation.

The triad lengths $l_i$ are not parameters, as in principle they are
evaluated for every set of quark positions in order to find the
potential in which the quarks move. Their average size does of course
affect the excitation energy of the string, and since this is
determined by solving for the motion of the quarks in the confining
(and to a lesser extent short-distance) potential, it only depends on
the quark masses and string tension.

In the case that the hyperfine contact spin-spin term in
Eq.~\ref{pot1} is set to zero, the lightest $L_q=0$
states have $M_{H_1} - M_B = 890$ MeV, giving a mass estimate of
$M_{H_1} = 1085 + 890 = 1975$ MeV. Here 1085 MeV is the spin-averaged
mass of the nucleon and $\Delta$ ground states. Note that this means
that all of the lowest-lying $H_1$ hybrid states in Table~\ref{tabqu}
have this mass. Furthermore, the states built on this adiabatic
surface with $L_q=1$ and $L_q=2$ have masses 2340 and 2620 MeV,
respectively, showing a considerable cost in energy for orbital
excitation of the quarks, comparable to that in the conventional
baryons. Similarly, the lightest radial excitation built on this
adiabatic surface has mass 2485 MeV, with a position between the
$L_q=0$ and $L_q=2$ states, as is also the case in conventional
baryons.

Including the hyperfine contact spin-spin term in Eq.~\ref{pot1}
lowers the mass of the quark-spin-1/2 $H_1$ hybrid states, which
coincide with the lightest $N$ hybrids as outlined in Sec.~\ref{qmno},
by 110 MeV to 1865 MeV, and raises the mass of the quark-spin-3/2
$H_1$ hybrid states, which coincide with the lightest $\Delta$
hybrids, by a similar amount.

These mass predictions depend on the form of the quark Hamiltonian
used here (Eq.~\ref{pot1}), which has been fit to the baryon
spectrum. The parameters which determine the adiabatic potential are
the string tension $b$ and the sum of the quark masses $\sum_i
M_i$. In order to conservatively estimate the error in the hybrid
masses due to uncertainties in these parameters, a variation of
$\delta b/b=\pm 10\%$, and $\delta \sum_i M_i/\sum_i M_i=+50\%$ have
been considered. These variations change the mass prediction for the
lightest hybrid by less than $\pm 100$ MeV.

These mass estimates are substantially higher than other mass
estimates in the literature, which are approximately $1.5$ GeV in the
bag model~\cite{bag} and $1.5\pm 0.15$ GeV in QCD sum
rules~\cite{QCDsr}.

There are two crucial assumptions that were made in the early work on
hybrid meson masses in the flux-tube model: the adiabatic motion of
quarks and the small-oscillations approximation for bead
motion~\cite{paton85}. It was later shown that when the adiabatic
approximation is lifted, the masses go up, and when the
small-oscillations approximation is lifted, the masses go
down~\cite{barnes95}. In the present study of hybrid baryons, the
adiabatic approximation has been partially lifted by the use of the
redefined adiabatic approximation. The small-oscillations
approximation has been fully lifted. The effect on the masses of
hybrid baryons when the various approximations are lifted is the same
as those found for hybrid mesons.

In the numerical simulation described above, with an identical Coulomb
interaction in conventional and hybrid baryons, the rms values
$\sqrt{\langle\rho^2\rangle}=\sqrt{\langle\lambda^2\rangle} = 2.12$,
$2.52$ GeV$^{-1}$ are obtained for the baryon and $H_1$ hybrid baryon
sizes, respectively.  The result
$\sqrt{\langle\rho^2\rangle}=\sqrt{\langle\lambda^2\rangle}$ is
expected since the spatial parts of the wave functions of the
low-lying states are totally symmetric under exchange symmetry. The
hybrid baryon is, therefore, 20\% larger in size than the conventional
baryon.

It is of physical interest to obtain an estimate of the effective
junction mass $\meff^{\infty}$ in Eq.~\ref{meff2}. Taking the average
values of $\sqrt{\langle\rho^2\rangle}$ and
$\sqrt{\langle\lambda^2\rangle}$ above with the quarks in an
equilateral triangle configuration, with $b=0.18$ GeV$^2$, and
using Eq.~\ref{Jacobi}, the junction mass is $\meff^{\infty} = 0.17$, $0.20$
GeV for the baryon and $H_1$ hybrid baryon, respectively. This
effective mass is made smaller by the center-of-mass corrections due
to the quark motion in Eq.~\ref{meff2}. This effective junction mass
found in the flux-tube model in the refined adiabatic approximation is
very different from the constituent gluon mass of $0.8$ GeV typically
employed in constituent gluon models~\cite{swanson98}, and partially
accounts for the higher excitation energy of the hybrid in the present
work.

\section{Discussion\label{discuss}}

It is interesting to compare results found here for hybrid baryons to
the predictions of the bag model~\cite{bag}. Out of all of the
flux-tube model states listed under $H_1^S$ and $H_1^A$ in
Table~\ref{tabqu}, only the $N^2 {\frac{1}{2}}^+$, and $N^2
{\frac{3}{2}}^+$ states have the same flavor, quark-spin $S$, total
angular momentum and parity as the low-lying hybrid baryons in the bag
model. However, restricting to experimentally measurable quantum
numbers (flavor, total angular momentum, and parity), only one
light-hybrid is different between the flux-tube and bag model
predictions. This $J^P=5/2^+$ state is flavor $\Delta$ in the
flux-tube model and flavor $N$ in the bag model, but is amongst the
higher-lying states in both models~\cite{bag}.

Looking at quantum numbers alone, the Roper resonance could be
regarded as a hybrid baryon candidate in both models. However, our
mass estimates do not support this identification, in contrast to some
bag model~\cite{bag} and QCD sum rule estimates~\cite{QCDsr}.  Both
our model and the bag model have seven low-lying hybrid baryons.

One of the disadvantages of hybrid baryons relative to hybrid mesons,
multi-quark states and glueballs, is that all possible baryon quantum
numbers can be attained by conventional baryons, so there are no
``exotic'' quantum numbers.  However, our low-lying $H_1^A$ hybrid
baryons are ``non-relativistic quark model exotic'', since these
states have quark-label exchange antisymmetric flux wavefunctions,
and so totally antisymmetric space $\times$ spin $\times$ flavor
wavefunctions, and states of this nature cannot be constructed in the
non-relativistic quark model.

The phenomenology of hybrid baryons has been reviewed recently
\cite{page02,barnes00} and so will not be discussed here.

\section{Conclusions\label{conc}}

Significant progress has been made towards building a realistic
flux-tube model of (hybrid) baryons.  The full multi-bead Hamiltonian
is constructed and it is demonstrated that the junction decouples from
the beads on the triads to a high degree of accuracy (with the exact
and decoupled lowest frequencies differing by at most 2\%), so that
the lowest-lying hybrid-baryon excitations in the flux-tube model can
be associated with the motion of the junction.  This simplifies the
description of the long distance properties of a low-lying (hybrid)
baryon to be that of three quarks and a junction, with the junction
connected to each of the quarks via a linear potential.  The parameter
dependence of the conventional and hybrid-baryon potential is {\it
predicted}, and can be used as an input in various theoretical
approaches.
 
The quantum numbers of the lightest hybrid baryons in the flux-tube
model are given in Table~\ref{tabqu}, and in the presence of the
expected spin-spin interactions between the quarks, the lightest
hybrid baryons are four nucleons with $J^P=1/2^+$, and $3/2^+$ and with a
mass of 1865$\pm$ 100 MeV.

\vspace{.5cm}
{\bf Acknowledgements}

Helpful discussions with T.~Barnes, T.~Cohen, N.~Isgur, G.~Karl,
R.~Lebed, A.V.~Nefediev, R.~Pack, and E.S.~Swanson are gratefully
acknowledged.  This work is supported by the U.S. Department of Energy under
Contracts DE-FG02-86ER40273 (SC) and W-7405-ENG-36 (PRP).

\appendix

\section{\label{junct} Action of the plaquette operator on the junction}

It is now shown that the junction can move and leave the string in the
ground state of its Y-shaped configuration in first order HLGT
perturbation theory, i.e. with the application of a single plaquette
operator.

In Fig.~\ref{lattice_plaq} the plaquette operator which operates on
the junction has the effect of producing a link which is indicated
with two arrows on it. The color structure of this link is ${\bf 3} \otimes
{\bf 3} = {\bf \bar{3}} \oplus {\bf 6}$. A link with color ${\bf 6}$ is not
allowed by conservation of color, because its neighboring link is
color ${\bf 3}$ flowing in the opposite direction, or color ${\bf
\bar{3}}$. However, it would appear that a link with color ${\bf
\bar{3}}$ is allowed. The following shows that this is the case.

\begin{figure}
\includegraphics[width=10cm,angle=0]{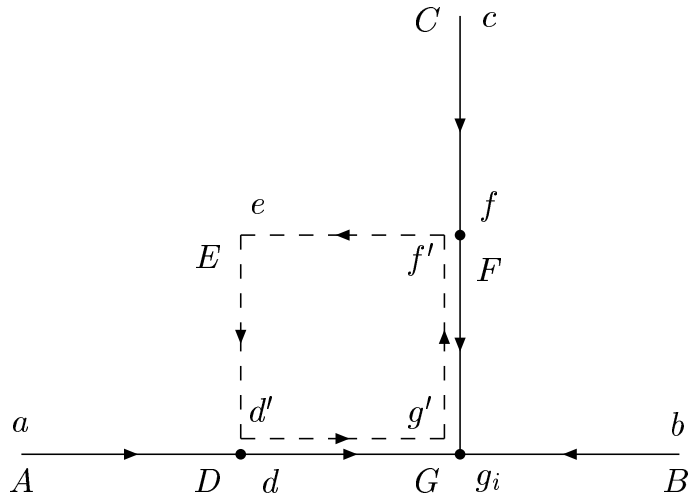}
\caption{\label{lattice_plaq_en} Enlarged view of the effect of the
plaquette operator on the junction in Fig.~\protect\ref{lattice_plaq}.}
\end{figure}

An enlargement of the junction region of Fig.~\ref{lattice_plaq} is
provided in Fig.~\ref{lattice_plaq_en}. The creation of a link from
spatial positions $A$ to $B$ with color projections $a$ and $b$
respectively, is denoted by $U^{ab}_{AB}$. Denoting by capital letters
the spatial  positions, and by lowercase letters the color projections,
the mathematical expression corresponding to the graph in 
Fig.~\ref{lattice_plaq_en} is
\begin{equation}\label{links}
\sum_{dfg_1g_2g_3f'ed'g'} \; \;
U^{ad}_{AD}\: U^{dg_1}_{DG}\: U^{bg_2}_{BG}\: U^{fg_3}_{FG}\: U^{cf}_{CF}\:
\epsilon_{g_1g_2g_3} \; \; \;
U^{f'e}_{FE}\: U^{ed'}_{ED}\: U^{d'g'}_{DG}\: U^{g'f'}_{GF},
\end{equation}
where the first term is the initial baryon state and the second term
is the plaquette operator. Now contract the two $DG$ links, and the
two $FG$ links, using Appendix B of ref.~\cite{kogut2}
\begin{eqnarray}
U^{dg_1}_{DG} U^{d'g'}_{DG} & = &
\frac{1}{2}\sum_{d''g''}\epsilon_{dd'd''}\epsilon_{g_1g'g''}U^{g''d''}_{GD} 
 + \nonumber\\
& & \hspace{-0.5cm}\frac{1}{4}\sum_{d''d'''g''g'''} (\delta_{dd''}
\delta_{d'd'''}+
\delta_{dd'''}\delta_{d'd''})\; (\delta_{g_1g''}\delta_{g'g'''}+
\delta_{g_1g'''}\delta_{g'g''})\; U_{DG}^{{\bf 6}\:\{ d''d'''\}\: \{ g''g'''
\} }\\
U^{fg_3}_{FG} U^{g'f'}_{GF} & = &
\frac{1}{3}\;\delta_{ff'} \delta_{g_3g'} + 
2\sum_{f''g''} \frac{\lambda^{f''}_{ff'}}{2}\frac{\lambda^{g''}_{g_3g'}}{2}\;
U_{FG}^{{\bf 8}\: f''g''},
\end{eqnarray}
where $U^{{\bf 6}}$ and $U^{{\bf 8}}$ refer to sextet and octet links
respectively, and $\lambda^a$ are the usual Gell-Mann SU(3) matrices.
It follows that Eq.~\ref{links} equals
%
\begin{equation}\label{linker}
-\frac{1}{3}\sum_{g''fedd'd''} \; 
U^{ad}_{AD}\: U^{bg''}_{BG}\: U^{g''d''}_{GD}\:U^{cf}_{CF}\: U^{fe}_{FE}
\: U^{ed'}_{ED}\:
\epsilon_{dd'd''},
\end{equation}
where the ${\bf 6}$ term does not contribute as mentioned above, and
the ${\bf 8}$ term is not displayed as it leads to a more complicated
topology. Eq.~\ref{linker} is a mathematical expression which gives a
graph similar to Fig.~\ref{lattice_plaq_en}, with the junction moved
from spatial position $G$ to position $D$, and the links in their
usual color triplet state. Hence the application of a single plaquette
operator can move the junction and leave the string in its ground
state, as promised.

\section{Flux-Tube Hamiltonian
\label{app1}}

The flux-tube Hamiltonian is first derived for the $\thetaj=120^o$
case, using the coordinate system defined in Eq.~\ref{xhat}. The goal
is to describe the position of each quark and bead and the junction
with respect to an alternative coordinate system with origin at the
center of mass $\rcm$ of the entire system.

In this coordinate system the junction is at ${\bf r}_J \equiv \br -
\rcm$, where $\br = (x,y,z)$. The quark positions are
\beqn\plabel{bdr2} 
{\bf r}_{i} \equiv l_i (-e^i_y,e^i_x,0)  - \rcm,
\eeqn
where the $\lpmb{e}^i$ are defined in Eq.~\ref{ees}. Bead $n$ on the
$i$-th triad is at position
\beqn 
\plabel{bdr1}   
{\bf r}^i_n \equiv
(-\frac{N_i+1-n}{N_i+1} e^i_y l_i + \frac{n}{N_i+1}x + \xi^i_n e^i_x,
 \frac{N_i+1-n}{N_i+1} e^i_x l_i + \frac{n}{N_i+1}y + \xi^i_n e^i_y,
\frac{n}{N_i+1} z + z^i_n) -\rcm.
\eeqn
Here $\xi^i_n$ is the displacement of bead $n$ along the direction
$\lpmb{e}^i$ which is perpendicular to the $i$-th triad in the QQQ
plane (see Fig.~\ref{bar_hybrid_beads}), and $z^i_n$ is its
displacement perpendicular to the QQQ plane. Note the bead
displacements are transverse to the triads on which they lie, and are
measured from the line connecting quark $i$ to the (in general
displaced) junction. Bead $n=1$ is placed next to the $i$-th quark,
and bead $N_i$ lies next to the junction.

Requiring that the center of mass is at the origin gives the
constraint
\beqn
{\bf 0} = m_J {\bf r}_J + m_b \sum_{i=1}^3\sum_{n=1}^{N_i} {\bf r}^i_n +  
\sum_{i=1}^3 M_i{\bf r}_{i},
\eeqn
which can be solved for $\rcm$. The Hamiltonian is
\beqn 
\plabel{hfluxraw}
H_{\mbox{\small flux}} \equiv
\frac{1}{2} m_J \dot{{\bf r}}_J^2 +
\frac{1}{2} m_b \sum_{i=1}^3\sum_{n=1}^{N_i} (\dot{{\bf r}}^i_n)^2  +
\frac{1}{2} \sum_{i=1}^3 M_i \dot{{\bf r}}_i^2 
+\sum_{i=1}^3\sum_{n=0}^{N_i} |{\bf r}^i_{n+1}-{\bf r}^i_{n}|,
\eeqn
where the first three terms form the kinetic energy. The last term is
the linear potential energy between neighboring beads, where
${\bf r}^i_{N_i+1}$ is defined as the position of the junction ${\bf r}_J$
and ${\bf r}^i_{0}$ is defined as the quark position ${\bf r}_{i}$.

The Hamiltonian is now simplified using the redefined adiabatic
approximation, where the distances between the quarks remain fixed,
concisely stated as $\dot{l}_i=0$. The small-oscillations
approximation is used to Taylor expand the potential energy in
Eq.~\ref{hfluxraw} to yield a function quadratic in the junction and
bead position coordinates.

The displacements of the beads within and out of the $QQQ$ plane,
given by $\xi^i_n$ and $z^i_n$, is expanded in terms of the amplitudes
of the $m$-th normal mode of the beads, $q_{\parallel m}^{i}$ and
$q_{\perp m}^{i}$, respectively, by
\beqn
\xi^i_n = \sum_{m=1}^{N_i} q_{\parallel m}^{i} \sin \frac{m n \pi}{N_i+1}
\hspace{1cm}
z^i_n = \sum_{m=1}^{N_i} q_{\perp m}^{i} \sin \frac{m n \pi}{N_i+1}.
\eeqn
The potential energy now becomes diagonal in these normal mode
amplitudes  $q_{\parallel m}^{i}$, $q_{\perp m}^{i}$ of the beads.
Carrying out the necessary algebra yields Eq.~\ref{ham1}.  As an
aside, it is perfectly meaningful in Eq.~\ref{ham1} to put
$N_1=N_2=N_3=0$, the case with no beads other than the junction.

For the $\thetaj>120^o$ case the definition of the coordinates in
Eqs.~\ref{bdr2}--\ref{bdr1} remains the same, except that the
$\lpmb{e}^i$ are given by Eq.~\ref{eet} and there are no beads on the
third triad.  Carrying out the necessary algebra yields the
Hamiltonian discussed in section \ref{gsec}.

\section{Quark label exchange symmetry for the $\thetaj = 120^o$ case in the 
small-oscillations approximation \label{app2}}

It is now shown, using the explicit definition of $\etahat_z$, that
both quark-label exchange symmetric and antisymmetric realizations of
$\etahat_z$ are possible.  First consider the possibility $\sigma_z =1$
in the definition Eq.~\ref{etahat} of $\etahat_z$, so that $\etahat_z$
points along $\brho\times\blambda$. By inserting the expressions for
$\lpmb{\rho}$ and $\lpmb{\lambda}$ in Eq.~\ref{rho} into
$\brho\times\blambda$, it can explicitly be verified that
$\brho\times\blambda$ is totally antisymmetric under exchange
symmetry. Hence, $\etahat_z$ with $\sigma_z =1$ is totally
antisymmetric under quark label exchange symmetry.

Another possibility is to choose $\etahat_z$ along the vector 
\beqn\plabel{etahatZ}
\left[\hat{\bf Z} \cdot {\brho}\times{\blambda} \right]{\brho}\times{\blambda},
\eeqn
where $\hat{\bf Z}$ is a space-fixed unit vector (not determined by
the quark positions). This is equivalent to choosing
\beqn\plabel{eps}
\sigma_z = \frac{\hat{\bf Z} \cdot {\brho}\times{\blambda}}{|\hat{\bf Z} \cdot {\brho}\times{\blambda}|}
\eeqn 
in Eq.~\ref{etahat}. This choice of $\sigma_z$ obviously yields a
totally symmetric $\etahat_z$.

Next consider the quark-label exchange symmetry of $\etahat_{-}$ and
$\etahat_{-}$ in the $\thetaj=120^o$ case in the small-oscillations
approximation, as defined in Eq.~\ref{etahat1}.  Consider the vectors
$\etahat_{\pm}^\prime$, defined by $\etahat_{\pm}=\sigma_{\pm}
\etahat_{\pm}^\prime$. Applying $P_{ij}$ exchanges $l_i
\leftrightarrow l_j$ in Eq.~\ref{etahat1}, as well as the labels $i$
and $j$ in $\hat{\bf x}$ and $\hat{\bf y}$ in Eq.~\ref{xhat}. Under
$P_{12}$, it is easy to see that $\etahat_{\pm}^\prime \rightarrow
-\etahat_{\pm}^\prime$.  Both $P_{13}$ and $P_{23}$ can be shown to
lead to $\etahat_{\pm}^\prime \rightarrow \pm \etahat_{\pm}^\prime$,
where the sign is dependent on the relative sizes of $l_1$, $l_2$, and
$l_3$.  For example, under $P_{23}$ one can show that this sign is
given by the sign of the expression
\beqn
\sqrt{s(l_1,l_2,l_3)}\pm (\frac{1}{l_2}
+\frac{1}{l_3}-\frac{2}{l_1}),
\eeqn
where $s(l_1,l_2,l_3)$ is defined in Eq.~\ref{sli}.

The fact that the $\etahat_{\pm}^\prime$ transform under label
exchange into themselves, up to a sign, follows from the observation
that the {\it ray} in which the $\etahat_{\pm}^\prime$ lie is the
physical line of oscillation of the junction in these vibrational
modes. Since label exchange does not change the physics, the
oscillation should still be in the same ray after label exchange, as
found. Since only the ray in which $\etahat_{\pm}$ lies is physical,
the possibility cannot be excluded that the $\etahat_{\pm}^\prime$ are
multiplied by a sign $\sigma_{\pm}$ when a standard choice of
eigenvectors is constructed, as in Eq.~\ref{etahat1}. It is possible
to show that a consistent set of sign conventions $\sigma_{\pm}$ can
be chosen such that the $\etahat_{\pm}$ are either totally symmetric
or antisymmetric under label exchange. Neither choice can be excluded.

It is also possible to show that the signs $\sigma_-$, $\sigma_+$ and
$\sigma_z$ are invariant under parity.


\begin{references}
%
\bibitem{Dytman} T.P.~Vrana, S.A.~Dytman, and T.S.~Lee,
Phys.\ Rept.\ {\bf 328}, 181 (2000).
%
\bibitem{Roper} Z.-P. Li, Phys. Rev. D {\bf 44}, 2841 (1991);
Z.-P.~Li, V.~Burkert, and Z.-J.~Li, Phys. Rev. D {\bf 46}, 70 (1992).
%
\bibitem{pageselectionrule} P.R. Page, {Phys. Lett. B} {\bf 402}, 183
(1997); F.E. Close and P.R. Page, {Nucl. Phys. B} {\bf 443}, 233 (1995); 
{Phys. Rev. D} {\bf 52}, 1706 (1995).
%
\bibitem{QCDsr} L.S. Kisslinger and Z.-P. Li, Phys. Rev. D {\bf 51},
 5986 (1995); L.S. Kisslinger, {Nucl. Phys. A} {\bf 629}, 30c (1998).
%
\bibitem{yan} C.-K. Chow, D. Pirjol, and T.-M. Yan, {Phys. Rev. D} {\bf 59},
056002 (1999). 
%
\bibitem{bag} T. Barnes and F.E. Close, {Phys. Lett. B} {\bf
123}, 89 (1983); E. Golowich,  E. Haqq, and 
G. Karl, {Phys. Rev. D} {\bf 28}, 160
(1983); C.E. Carlson and
T.H. Hansson, {Phys. Lett. B} {\bf 128}, 95 (1983);
I. Duck and E. Umland, {Phys. Lett. B} {\bf 128}, 221 (1983); 
C.E. Carlson, {Proc. of the $7^{th}$
Int. Conf. on the Structure of Baryons} (October 1995, Santa Fe, NM), 
 eds. B.F.~Gibson {\it et al.}, World Scientific, Singapore, 1996, p. 461.
%
\bibitem{seeCI} H. Hogaasen and J.M. Richard, Phys. Lett. B {\bf 134},
520 (1983).
%
\bibitem{capstick86} S. Capstick and N. Isgur, {Phys. Rev. D} {\bf 34}, 2809 
(1986).
%
\bibitem{RoperPC} S. Capstick, Phys. Rev. D {\bf 46}, 1965 (1992).
%
\bibitem{twopoles} R.A.~Arndt, J.M.~Ford, and L.D.~Roper,
Phys. Rev. D {\bf 32}, 1085 (1984).
%
\bibitem{CutkoskyP11} R.E. Cutkosky and S. Wang, Phys. Rev. D {\bf
42}, 235 (1990).
%
\bibitem{Krewald} J.~Speth, O.~Krehl, S.~Krewald, and C.~Hanhart,
Nucl.\ Phys.\ A {\bf 680}, 328 (2000).
%
\bibitem{paton85} N. Isgur and J. Paton, {Phys. Rev. D} {\bf 31}, 2910 (1985);
{Phys. Lett. B} {\bf 124}, 247 (1983); S. Capstick, S. Godfrey, N. Isgur, 
and J. Paton, {Phys. Lett. B} {\bf 175}, 457 (1986); I. Zakout and R. Sever,
{Z. Phys. C} {\bf 75}, 727 (1997).
%
\bibitem{deldar} G.S. Bali, {Nucl. Phys. Proc. Suppl.} {\bf 83}, 422 (2000);
S. Deldar, {Nucl. Phys. Proc. Suppl.} {\bf 83}, 440 (2000). 
%
\bibitem{bali} C. Michael, {Proc. of Quark Confinement and the
Hadron Spectrum III} (Confinement III), (Newport News, VA, 7-12 June 1998),
ed. N. Isgur, World Scientific, Singapore, p. 93;
%
\bibitem{Simonov} D.S.~Kuzmenko and Yu.A.~Simonov, hep-ph/0202277.
%
\bibitem{kharseev} D.E. Kharzeev, {Phys. Lett. B} {\bf 378}, 238 (1996);
B.Z. Kopeliovich, {Phys. Lett. B} {\bf 446}, 321 (1999); {\it ibid.} 
B {\bf 381}, 325 (1996); S.E. Vance, M. Gyulassy, and X.N. Wang, 
{Phys. Lett. B} {\bf 443}, 45 (1998). 
%
\bibitem{swanson98} E.S. Swanson and A.P. Szczepaniak, {Phys. Rev. D} 
{\bf 59}, 014035 (1999).  
%
\bibitem{morningstar97bag} K.J. Juge, J. Kuti, and C.J. Morningstar, 
{Nucl. Phys. Proc. Suppl.} {\bf 63}, 543 (1998). 
%
\bibitem{adia} T.J. Allen, M.G. Olsson, and S. Veseli, {Phys. Lett. B} 
{\bf 434}, 110 (1998).
%
\bibitem{page00} P.R. Page, {Proc. of The Physics of Excited Nucleons} 
(NSTAR2000), (Newport News, VA, 16-19 Feb. 2000), eds. V.D. Burkert, L. 
Elouadrhiri, J.J. Kelly, and R.C. Minehart, 
 World Scientific, Singapore, 2001, p. 171, nucl-th/0004053. 
%
\bibitem{CP1}
S.~Capstick and P.~R.~Page,
Phys.\ Rev.\ D {\bf 60}, 111501 (1999).
\bibitem{merlin85} J. Merlin and J. Paton, {Jour. Phys. G} {\bf 11}, 439 
(1985).
%
G.S. Bali, {Fizika B} {\bf 8}, 229 (1998);
%
T.T. Takahashi, H. Matsufuru, Y. Nemoto, and
H. Suganuma, {Phys. Rev. Lett.} {\bf 86}, 18 (2001);
%
{Proc. of Int. Symp. on Hadron and Nuclei} (Feb. 2001,
Seoul, Korea), published by Institute of Physics and Applied Phyics
(2001), ed. Dr. T.K. Choi, p. 341;
%
T.T. Takahashi, H. Matsufuru, Y. Nemoto, H. Suganuma, and T. Umeda,
{Nucl. Phys. Proc. Suppl.} {\bf 94}, 554 (2001).
%
\bibitem{luscher} M. Luscher, {Nucl. Phys. B} {\bf 180}, 317 (1981).
%
\bibitem{stephenson} C. Michael and P. Stephenson, {Phys. Rev. D} {\bf 50}, 
4634 (1994).
%
\bibitem{kalash} Yu.S.~Kalashnikova and A.V. Nefediev, {Phys. Atom. Nucl.} 
{\bf 60}, 1333 (1997); {Phys. Lett. B} {\bf 367}, 265 (1996).
%
\bibitem{morningstar97} K.J. Juge, J. Kuti, and C.J. Morningstar, 
{Nucl. Phys. Proc. Suppl.} {\bf 63}, 326 (1998); C.J. Morningstar, 
{Proc. of the $7^{th}$ Int. Conf. on Hadron Spectroscopy (Hadron'97)}, 
(Upton, NY, 25-30 Aug. 1997), eds. S.-U. Chung and H.J. Willutzki, 
American Institute of Physics, 1998, p. 136; C.J. Morningstar, {Proc. of 
Quark Confinement and the Hadron Spectrum III (Confinement III)}, 
(Newport News, VA, 7-12 June 1998), ed. N. Isgur, World Scientific, 
Singapore, p. 179, hep-lat/9809015. 
%
\bibitem{kogut} J. Kogut and L. Susskind, {Phys. Rev. D} {\bf 11}, 395 (1975).
%
\bibitem{capt} See, for example, S. Capstick, Proc. of Hadron
Spectroscopy and the Confinement Problem (Swansea, UK, 27 June - 8
July 1995), ed. D.V. Bugg, NATO ASI Series B: Vol. 353, Plenum Press,
1996, p. 311.
%
\bibitem{karsh} M. Hess, F. Karsch, E. Laermann, and I. Wetzorke, 
{Phys. Rev. D} {\bf 58}, 111502 (1999). 
%
\bibitem{barnes95} T. Barnes, F.E. Close, and E.S. Swanson, {Phys. Rev. D} 
{\bf  52}, 5242 (1995).
%
\bibitem{page02} P.R.~Page, Proc. of the 9$^{\rm th}$ Int. Conf. on
the Structure of Baryons (Baryons 2002), (3-8 March, 2002, Newport
News, VA), nucl-th/0204031.
%
\bibitem{barnes00} T. Barnes, {contribution to the COSY Workshop on Baryon 
Excitations} (May 2000, J\"{u}lich, Germany), nucl-th/0009011.
%
\bibitem{kogut2} J.~Kogut, D.K.~Sinclair and L.~Susskind, Nucl. Phys. B
{\bf 114}, 199 (1976).
%
\end{references}
\end{document}